\documentclass[prx,twocolumn,longbibliography,superscriptaddress,preprintnumbers]{revtex4-2}
\usepackage{amsmath}
\usepackage{amssymb}
\usepackage{bm}
\usepackage{braket}
\usepackage{amsthm}
\usepackage{diagbox}
\usepackage{setspace}
\usepackage{verbatim}
\usepackage{esint}
\usepackage{physics}
\usepackage{indentfirst}
\usepackage{graphicx}
\usepackage{float}
\usepackage{dsfont}
\usepackage{mathrsfs}
\usepackage{mathtools}
\usepackage{tikz-cd}
\usepackage{simplewick}
\usepackage{slashed}
\usepackage{mathtools}
\usepackage{hyperref}
\usepackage{slashed}
\usepackage{cancel}
\usepackage{xcolor}
\usepackage[normalem]{ulem}
\usepackage{youngtab}
\usepackage{ytableau}
\usepackage{enumerate}
\usepackage{svg}
\usepackage{hhline}

\renewcommand{\vec}[1]{\oldbm{#1}}

\renewcommand{\vec}[1]{\boldsymbol{#1}}

\def\bk{{\vec K}}
\def\bk{{\vec k}}

\def\bg{{\vec g}}

\def\bA{{\vec A}}

\def\bz{{\vec z}}

\def\ba{{\vec a}}
\def\bz{{\vec z}}
\def\bK{{\vec K}}
\def\bq{{\vec q}}
\def\bQ{{\vec Q}}

\def\bR{{\vec R}}
\def\bG{{\vec G}}

\def\bl{{\vec l}}

\def\bn{{\vec n}}

\def\br{{\vec r}}
\def\magt{\text{mag}}

\def\bnabla{{\boldsymbol \nabla}}
\def\P{{\mathcal{P}}}
\def\LLL{{\text{LLL}}}

\def\N{{\mathcal{N}}}

\def\LG{\text{(LG)}}
\def\mBloch{\text{mag}}
\def\naut{{(0)}}
\def\BZ{\text{BZ}}
\newcommand{\SU}{\mathrm{SU}}
\newcommand{\U}{\mathrm{U}}

\newcommand{\eqnref}[1]{Eq.\,\eqref{#1}}
\newcommand{\figref}[1]{Fig.\,\ref{#1}}
\newcommand{\tabref}[1]{Tab.\,\ref{#1}}
\newcommand{\ov}{\overline}

\makeatletter
\newcommand\RedeclareMathOperator{%
  \@ifstar{\def\rmo@s{m}\rmo@redeclare}{\def\rmo@s{o}\rmo@redeclare}%
}
\newcommand\rmo@redeclare[2]{%
  \begingroup \escapechar\m@ne\xdef\@gtempa{{\string#1}}\endgroup
  \expandafter\@ifundefined\@gtempa
     {\@latex@error{\noexpand#1undefined}\@ehc}%
     \relax
  \expandafter\rmo@declmathop\rmo@s{#1}{#2}}
\newcommand\rmo@declmathop[3]{%
  \DeclareRobustCommand{#2}{\qopname\newmcodes@#1{#3}}%
}
\@onlypreamble\RedeclareMathOperator
\makeatother
\RedeclareMathOperator{\Im}{Im}
\RedeclareMathOperator{\Re}{Re}

\makeatletter
\let\oldabs\abs
\def\abs{\@ifstar{\oldabs}{\oldabs*}}
\let\oldnorm\norm
\def\norm{\@ifstar{\oldnorm}{\oldnorm*}}
\makeatother

\begin{document}

\title{Exact Many-Body Ground States from Decomposition of Ideal Higher Chern Bands: Applications to Chirally Twisted Graphene Multilayers}
\author{Junkai Dong}
\email{junkaidong@g.harvard.edu}
\affiliation{Department of Physics, Harvard University, Cambridge, MA 02138, USA}
\author{Patrick J. Ledwith}
\affiliation{Department of Physics, Harvard University, Cambridge, MA 02138, USA}
\author{Eslam Khalaf}
\affiliation{Department of Physics, The University of Texas at Austin, TX 78712, USA}
\author{Jong Yeon Lee}
\affiliation{Kavli Institute for Theoretical Physics, University of California, Santa Barbara, CA 93106, USA}
\author{Ashvin Vishwanath}
\affiliation{Department of Physics, Harvard University, Cambridge, MA 02138, USA}
\date{\today}
\begin{abstract}
Motivated by the higher Chern bands of twisted graphene multilayers, we consider flat  bands with arbitrary Chern number $C$ with ideal quantum geometry. While $C>1$ bands differ from Landau levels, we show that these bands host exact fractional Chern insulator (FCI) ground states for short range interactions. We show how to decompose ideal higher Chern bands into separate ideal bands with Chern number $1$ that are intertwined through translation and rotation symmetry. The decomposed bands admit an $\SU(C)$ action that combines real space and momentum space translations. Remarkably, they also allow for analytic construction of exact many-body ground states, such as generalized quantum Hall ferromagnets and FCIs, including flavor-singlet Halperin states and Laughlin ferromagnets in the limit of short-range interactions. In this limit, the $\SU(C)$ action is promoted to a symmetry on the ground state subspace. While flavor singlet states are translation symmetric, the flavor ferromagnets correspond to translation broken states
and admit charged skyrmion excitations corresponding to a spatially varying density wave pattern. We confirm our analytic predictions with numerical simulations of ideal bands of twisted chiral multilayers of graphene, and discuss consequences for experimentally accessible systems such as monolayer graphene twisted relative to a Bernal bilayer.
\end{abstract}
\maketitle

\section{Introduction}

The discovery of correlated states in twisted bilayer graphene (TBG) has inspired interest in the study of topological flat bands~\cite{Cao_2019,Yankowitz_2019,Lu_2019,Stepanov_2020,Cao_2021,Liu_2021}.
The interplay of band topology and strong interactions in TBG makes it an ideal candidate to realize strongly correlated topological physics. 
A simplified ``chiral" model\cite{tarnopolskyOriginMagicAngles2019}, where intrasublattice moir\'e tunneling is ignored, yields exactly flat topological bands at the magic angle, thereby capturing the most essential features of the system. The chiral model hosts analytically-attainable wavefunctions~\cite{tarnopolskyOriginMagicAngles2019} that are equivalent to those of the LLL in an inhomogeneous magnetic field~\cite{ledwithFractionalChernInsulator2020a}. Its mathematical properties and mysteries have turned the chiral model into a sub-field of study of its own right~\cite{tarnopolskyOriginMagicAngles2019,khalafMagicAngleHierarchy2019,ledwithFractionalChernInsulator2020a,wangChiralApproximationTwisted2021a,popovHiddenWaveFunction2020,renWKBEstimateBilayer2021a,naumisReductionTwistedBilayer2021a,navarro-labastidaWhyFirstMagicangle2022,navarro-labastidaHiddenConnectionTwisted2022,beckerSpectralCharacterizationMagic2020,beckerIntegrabilityChiralModel2022,beckerFineStructureFlat2022,shefferChiralMagicAngleTwisted2021a}.

At the same time, it was shown \cite{ledwithFractionalChernInsulator2020a} that the chiral model satisfies a quantum-geometric\cite{jacksonGeometricStabilityTopological2015,claassenPositionMomentumDualityFractional2015a,leeBandStructureEngineering2017,meraEngineeringGeometricallyFlat2021,meraKahlerGeometryChern2021,ozawaRelationsTopologyQuantum2021,zhangRevealingChernNumber2021,varjas_topological_2022,royBandGeometryFractional2014,wangExactLandauLevel2021b} identity known as the ``trace condition".
It was recently shown that the trace condition enables intra-band vortex attachment\cite{ledwithFamilyIdealChern2022,vortexability}, which has strong implications for interacting physics. For example, it directly enables the construction of FQHE-like trial states which are {\it exact} ground states of the system under short-range repulsive interactions. For TBG, this led to the analytic prediction\cite{ledwithFractionalChernInsulator2020a} of fractional Chern insulators (FCI)\cite{liuRecentDevelopmentsFractional2022,parameswaranFractionalQuantumHall2013,BergholtzReview2013,neupertFractionalQuantumHall2011,shengFractionalQuantumHall2011,regnaultFractionalChernInsulator2011,qi_generic_2011,parameswaranFractionalChernInsulators2012,wuBlochModelWave2013,kourtisFractionalChernInsulators2014}, states that host the fractional quantum Hall effect at fractional filling of a Chern band first observed\cite{spantonObservationFractionalChern2018a} in Hofstadter bands \cite{mollerFractionalChernInsulators2015,andrewsStabilityFractionalChern2018,andrewsFractionalQuantumHall2020a,andrewsStabilityPhaseTransitions2021,bauerFractionalChernInsulators2022}.  FCIs in TBG were simultaneously predicted in numerical works~\cite{abouelkomsanParticleHoleDualityEmergent2020a,repellin_chern_2020,wilhelmInterplayFractionalChern2021a} and were recently experimentally observed~\cite{xieFractionalChernInsulators2021a} in a small magnetic field that effectively restored the ideal chiral limit \cite{parkerFieldtunedZerofieldFractional2021}. We will refer to trace condition satisfying bands as \emph{ideal} bands; they have also been referred to as ``vortexable" \cite{vortexability}. The trace condition is also related to momentum space holomorphicity \cite{claassenPositionMomentumDualityFractional2015a,Lee2018,meraKahlerGeometryChern2021,meraEngineeringGeometricallyFlat2021,wangExactLandauLevel2021b,vortexability}, a powerful analytic tool.

Multilayer twisted graphene systems consisting of $n$ chirally stacked\cite{zhangSpontaneousQuantumHall2011,shiElectronicPhaseSeparation2020,geisenhofQuantumAnomalousHall2021} graphene layers, i.e. AB, ABC, etc., stacked with a single twist on top of $m$ chirally stacked layers are a natural extension of TBG. Experimentally, twisted double bilayer graphene, $(n,m) = (2,2)$~\cite{bibi1,bibi2,bibi3,bibi4,bibi5,bibi6,bibi7,bibi8Liu}, and twisted mono-bilayer graphene $(n,m) = (1,2)$~\cite{monobi1,monobi2,monobi3,monobi4,polshynElectricalSwitchingMagnetic2020,TCDW,monobiLi_STM,monobiTong_STM}, have been fabricated and a nearly quantized anomalous Hall effect has repeatedly been observed with $C>1$~\cite{bibi1,monobi1,monobi2,bibi7}. This has generated extensive theoretical interest in studying topological physics in these systems at different fillings~\cite{zhangTwistedBilayerGraphene2019a,liuQuantumValleyHall2019,haddadiMoireFlatBands2020,zhangChiralDecompositionTwisted2020,maMoirBackslashFlat2020,chebroluFlatBandsTwisted2019,Liu_bibiFCI,ledwithFamilyIdealChern2022,wangHierarchyIdealFlatbands2021,zhangSpinPolarizedNematic,LiangMoireBandStructure,luo2022strain,MaTwistedTrilayer,WangPhaseDiagramBibi}. Furthermore, symmetry broken Chern insulators have been found experimentally at half filling~\cite{TCDW}. A variety of these states was recently identified numerically together with signs of an approximate $\SU(2)$ symmetry~\cite{Wilhelm2204}, although its origin has not been understood. FCIs have also been studied in these systems \cite{Liu_bibiFCI,wangExactLandauLevel2021b}, and other systems with higher Chern bands \cite{wuBlochModelWave2013,behrmannModelFractionalChern2016,wangFractionalQuantumHall2012,liuFractionalChernInsulators2012,trescherFlatBandsHigher2012,yangTopologicalFlatBand2012,sterdyniakSeriesAbelianNonAbelian2013,andrewsStabilityFractionalChern2018,andrewsStabilityPhaseTransitions2021}, though analytic progress has largely been limited to toy models. These advances motivate the need for a systematic understanding of correlation phenomena in higher Chern bands, including both symmetry-broken and fractional Chern insulators. 

Chiral models have been constructed for the twisted chiral multilayers in Refs.~\cite{ledwithFamilyIdealChern2022,wangHierarchyIdealFlatbands2021} through neglecting certain interlayer tunneling terms.  Chiral twisted multilayer graphene yields flat and ideal higher Chern bands that are analytically tractable when tuned to the same magic angle as chiral TBG.
By combining ideality, which allows the construction of trial wavefunctions for topologically ordered states, with higher Chern number in an experimentally feasible system, these models provide a novel and unexplored platform to study and realize interacting topological phases. Ref. \cite{wangHierarchyIdealFlatbands2021} numerically found model FCI states with particle entanglement suggestive of particular ``color-singlet" \cite{wuBlochModelWave2013} Halperin states, though a general analytic understanding of the range of possible states in such a system remains lacking. \par 

One approach to understanding correlated states in a $C$\,$>$\,$1$ band is to employ the hybrid Wannier functions to decompose the band into $C$ Chern 1 bands by enlarging the unit cell \cite{barkeshliTopologicalNematicStates2012a}. However, this approach does not respect the ideality condition: an ideal Chern $C>1$ band generally decomposes into a set of $C = 1$ non-ideal bands such that analytic techniques cannot be applied to the decomposed basis. Furthermore, this decomposition assumes broken translation symmetry in a specific direction from the outset, making it difficult to understand translation-unbroken states like the FCIs observed in Ref.~\cite{wangHierarchyIdealFlatbands2021}. This naturally leads to the question: is it possible to decompose an ideal $C>1$ band into ideal $C=1$ bands and in a way that captures both translationally symmetric and translation-breaking states? A partial answer to this question was provided in Ref.~\cite{ledwithFamilyIdealChern2022} which showed that it is impossible to decompose a generic ideal $C>1$ bands in terms of {\it orthogonal} and ideal $C = 1$ bands.

Here, we show that by lifting the orthogonality constraint, it is possible to decompose a generic ideal $C>1$ into $C$ ideal Chern $1$ bands. From this decomposition, we reveal a hidden {\it non-unitary} $\SU(C)$ action among the decomposed bands combining real space and momentum space translations. We note that although this $\SU(C)$ action is generally non-unitary and is not a symmetry of the Hamiltonian, it {\em is} often a symmetry of the ground state manifold for short-range repulsive interactions, enabling us to make sharp predictions regarding the many-body ground state at partial filling. Furthermore, the $\SU(C)$ structure is reminiscent of the structure of multi-component Landau levels \cite{wuBlochModelWave2013}, allowing us to interpret these ground states in terms of more familiar correlated states in multi-component quantum Hall systems. It should be emphasized that the states we obtain are physically distinct and have non-trivial translation symmetry breaking patterns visible through real-space charge density, compared to multi-component Landau levels which have uniform charge density.

By employing this decomposition, we analytically construct ground states at a variety of fillings of ideal Chern $C>1$ bands and make concrete predictions on their realization.
First, we identify a ground state manifold of charge density waves (CDWs) at filling $1/C$ with emergent $\SU(C)$ symmetry. Such states can be interpreted as generalized quantum Hall ferromagnets in the decomposed basis. An immediate consequence of this identification is the existence of charged skyrmion textures which correspond to a characteristic winding pattern of the CDW order parameter (cf.~Fig.~\ref{fig:results}). Such a winding pattern can be readily observed with local charge probes such as STM, which has already been employed in these systems~\cite{Liu_STM,Verdu_STM,Crommie_STM1,Crommie_STM2,samajdar_STM,monobiLi_STM,monobiTong_STM}. Second, we characterize the structure of translation symmetric fractional Chern insulator states at fillings $1/(2Cs+1)$, for each positive integer $s$, by establishing a direct analogy with flavor-singlet Halperin states. 
Finally, we discuss a manifold of translation-breaking Laughlin states that appear at low fillings $1/{C(2s+1)}$ where fractionalization and topological order coexists with CDW order. All our results are verified by numerical exact diagonalization (ED) on the chiral model for twisted mono-bilayer graphene. These results enable us to make experimental predictions for graphene multilayers.

The paper is organized as follows. We begin with an overview of the central theoretical results as well as some consequences for experiment in Sec.~\ref{sec:results}. Next, we review the physical model for chiral graphene multilayers in Sec.~\ref{sec:graphene} and quantum geometry techniques to understand the single particle physics in Sec.~\ref{sec:quantumgeometry}. We present the central technical result, the decomposition of an ideal higher Chern band, in Sec.~\ref{decomp} for $C=2$. We study quantum ferromagnets, which manifest as topological CDWs, and their associated skyrmion description at half filling for a spinless $C=2$ band in Sec.~\ref{sec:cdw} and a spinful $C=2$ band in Sec.~\ref{sec:spin}. We discuss the possible fractional Chern insulators, both translation symmetric and translation broken, in Sec.~\ref{sec:FCIs}. We generalize the physical implications to ideal bands with $C>1$ in Sec.~\ref{sec:higherC}. We conclude with some future questions in Sec.~\ref{sec:conclusion}.

\section{Summary of Results}\label{sec:results}
We begin by briefly summarizing our results and discussing their main implications. By now, a full understanding of ideal $C = 1$ bands has been achieved: they have been related to the LLL of a Dirac particle in a magnetic field~\cite{wangExactLandauLevel2021b,ledwithFractionalChernInsulator2020a}, which enables the construction of exact many body Laughlin-like ground states for short range interactions~\cite{vortexability}. Our main technical achievement in this work is the generalization of such understanding for any ideal higher Chern band. In particular, we provide a general explicit construction to decompose any ideal higher Chern band into ideal but non-orthogonal Chern 1 bands with a non-unitary $\SU(C)$ action.
Compared to the hybrid Wannier decomposition~\cite{barkeshliTopologicalNematicStates2012a}, our procedure has the distinct advantages of preserving the ideal band geometry and allowing access to translation broken states in all possible directions in the same basis. The preservation of ideal band geometry means we leverage the analytic knowledge of ideal $C=1$ bands to analytically study interacting phases -- in particular, FCIs -- in $C>1$ bands.
\begin{figure*}
    \centering
    \includegraphics[width=0.85\linewidth]{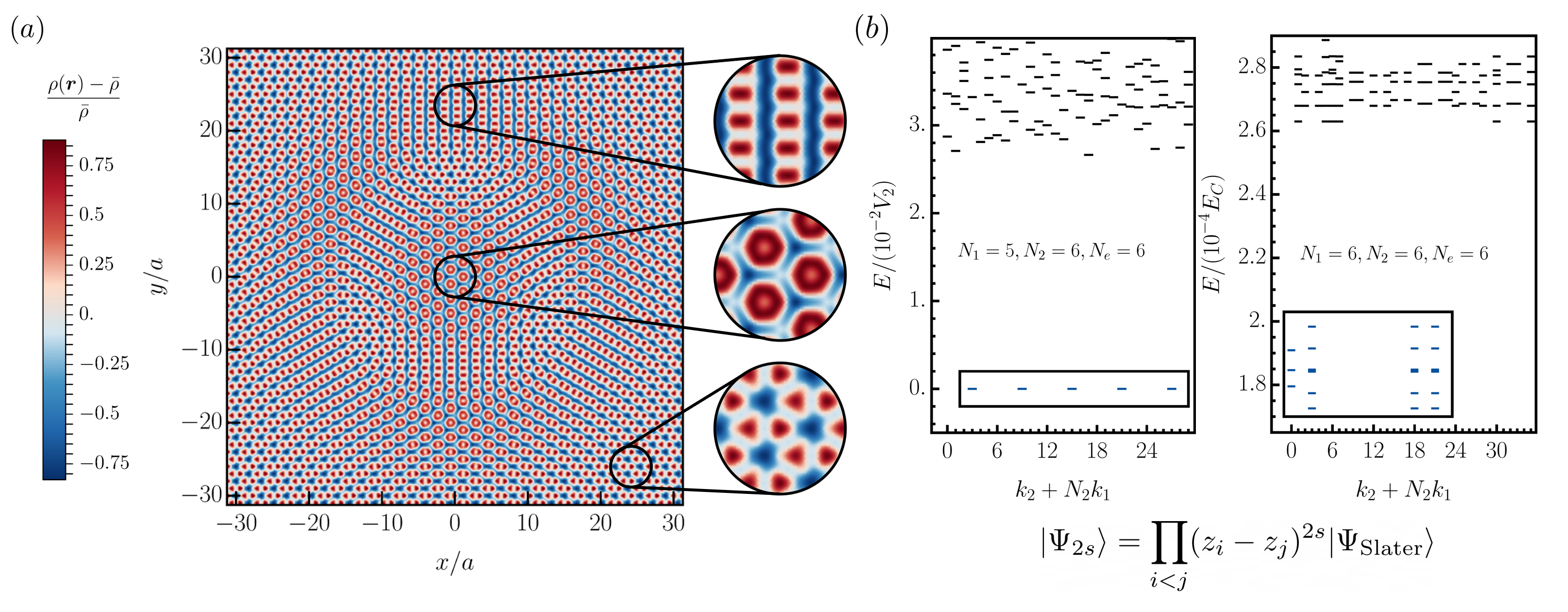}
    \caption{Main results of our paper. (a) Plots of the density profile of a CDW-skyrmion that arises out of a manifold of topological charge density waves from the ideal $C=2$ band in chiral twisted monolayer-bilayer graphene. Windings of the order parameter corresponds to windings of the translation breaking pattern. We zoom into different regions to highlight different regions that correspond to particular topological charge density waves that repeat in a $2 \times 2$ unit cell. (b) shows many-body spectrum from ED at different fillings for the ideal $C=2$ band in chiral twisted monolayer-bilayer graphene: boxed energy levels correspond to the ground state manifold of model FCIs. The first panel is the translation invariant $(332)$ Halperin state at $\nu=1/5$ with the interaction $V_2\delta''(\hat{\br}_i-\hat{\br}_j)$ and the expected ground state degeneracy of $5$. The second panel is the translation broken Laughlin state at $\nu=1/6$ under a screened Coulomb interaction with a ground state degeneracy of $21$ that combines the Laughlin degeneracy of $3$ with the generalized-ferromagnetic degeneracy of $N_e + 1 = 7$. Both of these collections of states originate from the depicted analytic vortex attachment construction summarized in Sec.~\ref{sec:quantumgeometry}.}
    \label{fig:results}
    \end{figure*}\par
The decomposition provides analytic many-body states that are the {\it exact} ground states at certain fractional fillings for short range interactions. For example, for a spinless band with Chern number $C$ these ground states include topological CDWs at $1/C$ filling. At lower fillings, we construct Halperin-like translationally symmetric FCIs at $\nu = 1/(2Cs+1)$ and Laughlin-like translation broken FCIs at $\nu= 1/C(2s+1)$ for any positive integer $s$. 

\begin{figure}
    \centering
    \includegraphics[width=0.85\linewidth]{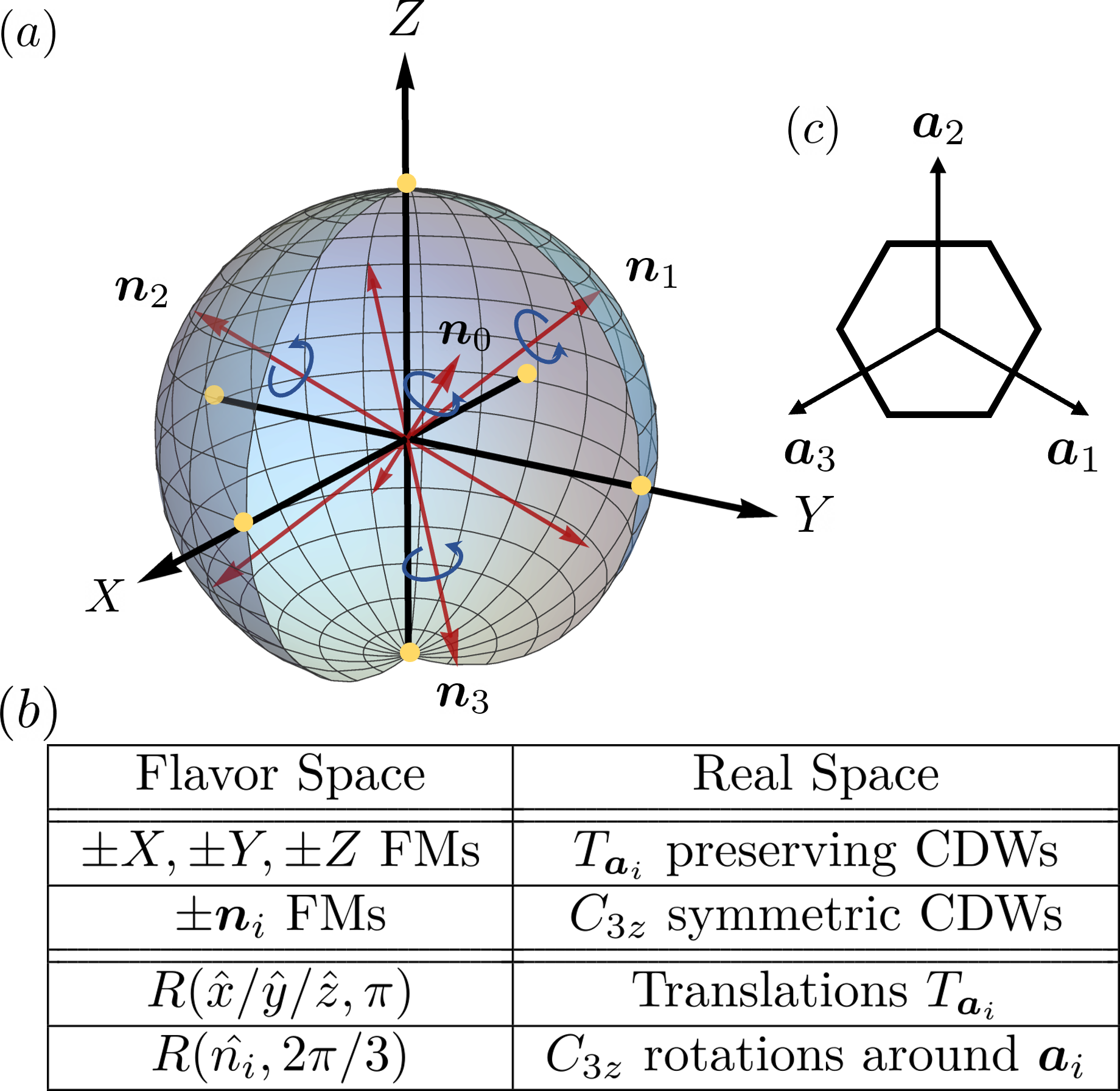}
    \caption{Sphere of flavor ferromagnets (FMs) at half filling for a $C_{3z}$ symmetric ideal band with $C=2$. Each point on the sphere correspond to a topological CDW with unit cell $2\ba_1\times 2\ba_2$. (a) We show the order parameter sphere of ferromagnets: red arrows correspond to $C_{3z}$ symmetric CDWs (second and third insets of \figref{fig:results}(a)) and yellow dots correspond to CDWs that preserve certain translation symmetries $T_{\ba_i}$ (first inset of \figref{fig:results}(a)). $C_{3z}$ rotations in real space correspond to rotations around red arrows in order parameter space. Table (b) explains the representation of special CDWs as special ferromagnets on the sphere and crystalline symmetries as rotations. (c) We show the real space unit cell and the lattice vectors.}
    \label{fig:ferrosphere}
    \end{figure}

The topological CDWs are best understood as a manifold of generalized quantum Hall ferromagnets (FM), where translations act as pseudospin rotations on the ferromagnets. The pseudospin ferromagnets are generically translation symmetric only in a $C \times C$ unit cell. Crystalline symmetries, both translations and discrete rotations, emerge as rotations in the order parameter manifold. For $C=2$, this manifold is a sphere and different topological CDW states may be understood as different pseudospin directions, pictorially represented in Fig.~\ref{fig:ferrosphere}(a) and  summarized in \tabref{fig:ferrosphere}(b).
The $\pm X, \pm Z, \pm Y$ axes of the pseudospin sphere correspond to states that preserve the translations along $\ba_{1,2,3}$, where $\ba_{1,2}$ are primitive lattice vectors and $\ba_{3} = -\ba_1 - \ba_2$.  For rotational symmetry, we focus on the three-fold $C_{3z}$ case which is relevant for graphene moir\`e systems, and other rotational symmetries may be understood similarly. We find that $C_{3z}$ acts as a $120^\circ$ rotation around a particular $\bn_0$ axis of the pseudospin sphere. There are also rotations $C_{3z}^{(i)} = T_{\ba_i} C_{3z}T_{-\ba_i}$ around lattice vectors $\ba_i$ and corresponding invariant axes $\bn_i$; these are distinct operations because we have broken translations.

We construct exact model FCIs with the vortex attachment procedure \cite{ledwithFamilyIdealChern2022,vortexability} reviewed in Sec.~\ref{sec:quantumgeometry}. By attaching vortices to the fully filled Chern $C$ band, we arrive at an explicit first quantized wavefunction for the translationally-symmetric Halperin states. If we instead choose a $C=1$ topological charge density wave as a parent state we obtain the translation broken Laughlin states. We numerically verify these predictions in exact diagonalization for the $C\,{=}\,2$ band of chiral twisted monolayer-bilayer graphene, see Fig. \ref{fig:results}(b). 

Finally, based on our analysis in the ideal limit, we make concrete predictions for the ground states and excitations of twisted mono-bilayer graphene at several fillings as follows. Note, here we allow for valley and spin degeneracy, hence the filling $\nu$ ranges from $0\leq \nu\leq 4$. We will focus on states doped on top of a $\nu=3$ spin and valley polarized state such that there is a single empty $C=2$ band left over.\begin{enumerate}
    \item At $\nu=3+1/2$, we expect that the ground state will be a continuous manifold of spin polarized topological CDWs, or generalized quantum Hall ferromagnets. The observed topological CDWs in Ref.~\cite{TCDW} are special points in a continuous manifold of generalized quantum Hall ferromagnets. Our main prediction is that there will be associated \textit{charged} skyrmion textures with a particular winding of translation breaking patterns illustrated in Fig.~\ref{fig:results}(a). These textures can be directly probed by STM.\par
    \item At $\nu=3+1/5$, we expect that the ground state will be a translationally invariant Halperin $(332)$ state: see first panel of Fig.~\ref{fig:results}(b) for numerical spectrum. In particular, the Hall conductivity (or the slope of the gapped feature on the Landau fan diagram) will be $2/5$.\par
    \item At $\nu=3+1/6$, we expect that the ground state will be translation breaking ferromagnetic Laughlin states: see second panel of Fig.~\ref{fig:results}(b) for numerical spectrum. These states will in general be translation breaking striped phases with $1/3$ Hall conductivity.\par
\end{enumerate}\par
We end our summary of results with some cautionary comments on application to experiments.
First, realistic systems will not be in the chiral limit, which means that the higher Chern band will not have exactly ideal quantum geometry, and may be dispersive. However, the band geometry and flatness  could be improved by external perturbations such as out of plane electric and magnetic fields as believed to occur in magic angle graphene  \cite{parkerFieldtunedZerofieldFractional2021}, bringing the system closer to the ideal limit. Second, in most of our discussions, we have assumed a spinless band for simplicity; exotic magnetic orders at half filling such as the tetrahedral antiferromagnetic (TAF) order exist in spinful bands~\cite{Wilhelm2204,TAF1,TAF2,TAF3,TAF4,TAF5,TAF6}. Remarkably, after adding spin we find that the TAF states lie in the same $\mathbb{CP}^3$ manifold as the previously mentioned charge density waves. Furthermore, a  magnetic field may be used to favor the spin-polarized submanifold that hosts charge-density skyrmions.
A final caveat is that typical twisted graphene systems contain twist angle disorder which leads to large inhomogeneous heterostrain. Strain greatly increases single particle dispersion and pushes the band away from ideality. 
We leave a detailed quantitative study of the realistic system that addresses the above issues  to future work.

\section{Chiral Twisted Graphene Multilayers}\label{sec:graphene}

In this section, we review the chiral twisted graphene multilayers: they give rise to ideal higher Chern bands and closely describe experimentally relevant systems. We use them as our primary numerical example. Experimentally, these graphene multilayers have been fabricated both as twisted monolayer-bilayer (mono-bi)~\cite{monobi1,monobi2,monobi3,monobi4} and twisted bilayer-bilayer (bi-bi)~\cite{bibi1,bibi2,bibi3,bibi4,bibi5} graphene systems. The following discussion will closely follow recent works which introduced these models~\cite{ledwithFamilyIdealChern2022,wangHierarchyIdealFlatbands2021}.  

We consider $n+m$ graphene layers. The first $n$ layers are untwisted and chirally stacked on each other: chiral stacking means that the successive layers have Bernal stacking AB or BA. These $n$ chirally stacked layers are then twisted by the magic angle $\theta$ of chiral TBG, and put on top of another $m$ layers of chirally stacked graphene. The mono-bilayer system corresponds to $n$\,=\,$1,m$\,=\,$2$ and the bilayer-bilayer system corresponds to $n$\,=\,$2,m$\,=\,$2$.

For concision, consider electrons of a specific valley and spin flavor. 
We make the approximation that the electron can only tunnel between neighboring layers, and that the interlayer tunneling is on-site (this neglects e.g. trigonal warping terms \cite{jungAccurateTightbindingModels2014}). The Hamiltonian can be written as
\begin{equation}\label{eq:multilayer_ham}
    H=\left(\begin{array}{cc}
        h_{m,\sigma} & T_M \\
        T_{M}^\dagger & h_{n,\sigma'}
    \end{array}\right)
\end{equation}
where $h_{n,\sigma}$ describes the $n$ untwisted chirally stacked graphene layers with chirality $\sigma = \pm$ depending on whether the layers have AB ($+1$) or BA ($-1$) stacking. $T_M$ only tunnels between layers $n$ and $n+1$ and is identical to the moir\'e tunneling of chiral TBG, where intra-sublattice tunneling is switched off. Its explicit dimensionless form is given by
\begin{equation}
    T_M = \alpha \left(\begin{array}{cc}
        0 & U(\br) \\
        U^*(-\br) & 0
    \end{array}\right)
    \label{eq:moiretunneling}
\end{equation}
where $U_1(\br) = \sum_{n=0}^2 e^{\frac{2\pi i n}{3} - i \bq_n \cdot \br}$, $\bq_n = C_3^n (0,-1)^T$, and $C_3$ is a $120^\circ$ rotation. 
The parameter $\alpha = w_1/vk_\theta$ is a dimensionless tunneling strength, where $w_1 \approx 110$ meV and the angle-dependent kinetic energy scale is $v k_\theta = v\abs{\bK^{\mathrm{top}} - \bK^{\mathrm{bot}}}$, where $v$ is the Fermi velocity of the Dirac cone and $\bK^{\mathrm{top,bot}}$ are the graphene $\bK$-points for the top and bottom stacks respectively.  The chiral TBG magic angle corresponds to $\alpha \approx 0.586$ \cite{tarnopolskyOriginMagicAngles2019}.

The form of $h_{n,\sigma}$ is tri-diagonal:
\begin{equation}
    h_{n,\pm}=\left(\begin{array}{cccc}
        -i\sigma\cdot\nabla & T_{\pm} & 0 &\dots \\
        T_{\pm}^\dagger & -i\sigma\cdot\nabla & T_{\pm} & \dots\\
        0 & T_{\pm}^\dagger & -i\sigma\cdot\nabla &\dots \\
        \dots &\dots &\dots &\ddots
    \end{array}\right)
\end{equation}
The diagonal terms describe the Dirac cones of each layer, and the off diagonal terms describe the tunneling due to two kinds of Bernal stacking: $T_\pm=\beta \frac{\sigma_x\pm i\sigma_y}{2}$.

Most importantly, the model hosts zero energy flat bands at particular values of $\alpha$ identical to the magic angles for chiral TBG. These zero energy bands are closely related to the chiral symmetry of the model: $\{H,\sigma_z\}\,{=}\,0$. The chiral symmetry arises because the model is purely off diagonal in the sublattice basis:
\begin{equation}
    H=\left(\begin{array}{cc}
        0 & \mathcal{D}^\dagger \\
        \mathcal{D} & 0
    \end{array}\right)_{\text{AB}},
    \label{eq:chiral_ham_form}
\end{equation}
where the subscript AB denotes that the matrix is block diagonalized in its sublattice. Thus, we may choose zero energy eigenstates of $H$ to be eigenstates of $\sigma_z$. At the magic angle $\alpha\,{=}\,0.586$, two zero energy bands labeled by the sublattice index $A$ and $B$ emerge independent of the value of $\beta$~\cite{ledwithFamilyIdealChern2022,wangHierarchyIdealFlatbands2021}. These bands carry Chern numbers, whose signs depend on the stacking chiralities $(\sigma,\sigma')$ and whose values depend on ($n$,$m$).

\begin{figure}
    \centering
    \includegraphics[width=0.75\linewidth]{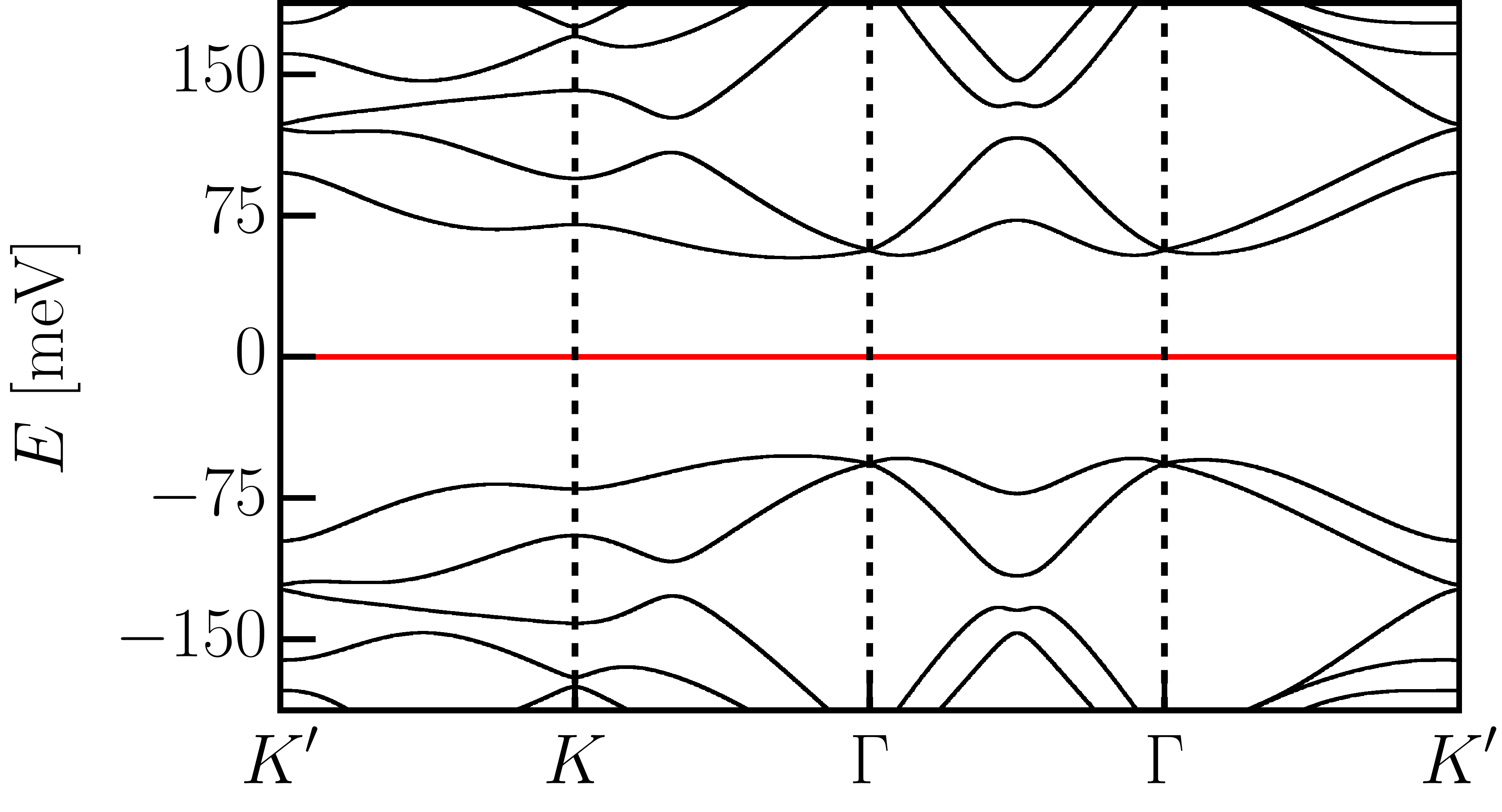}
    \caption{Band structure of twisted monolayer-bilayer graphene. The higher Chern band is highlighted in red. We use the chiral model \eqref{eq:multilayer_ham} with $\alpha=0.586$ and $\beta=2$; the two Chern bands are degenerate and both of them have zero energy.}
    \label{fig:band}
\end{figure}

For example, when both upper $n$ layers and lower $m$ layers are AB-stacked with $\sigma=\sigma'=1$, the Chern number for the $A$-sublattice polarized flatband (zero-energy) is $C_A\,{=}\,n$, while for the $B$-sublattice polarized flatband $C_B\,{=}\,{-}m$. In particular, both mono-bilayer and bilayer-bilayer systems give rise to Chern bands with $|C|=2$ for this stacking orientation. For an analytic derivation of the flat band wavefunctions $\psi_\bk(\br)$ for any stacking $(\sigma,\sigma')$, in terms of the $n=m=1$ chiral TBG wavefunctions, see Ref. \cite{ledwithFamilyIdealChern2022}. We numerically plot the band structure of mono-bi with the chiral model in Fig.~\ref{fig:band}; the higher Chern band is highlighted in red.

In the next section we will review ideal band geometry and show that the zero modes (flat bands) of magic-angle chiral graphene multilayers in \eqnref{eq:chiral_ham_form} attain the ideal geometry. Crucially, the zero mode operator $\cal{D}$ only depend on coordinates $x,y$ and the antiholomorphic derivative $\ov{\partial} = \frac{1}{2}(\partial_x + i \partial_y)$: ${\cal D}(\br,\partial_x,\partial_y)={\cal D}(\br,\ov{\partial})$. This means that  
\begin{equation}
    [{\cal D},z] = 0, \quad \text{where } z = x+iy.
    \label{eq:ideal_band_origin}
\end{equation}
This property of $\cal{D}$ will be equivalent to the ``vortex-attachment'' description of ideal quantum geometry.

\section{Quantum geometry of Ideal Chern bands}\label{sec:quantumgeometry}

In this section we review our characterization of ideal band geometry via the equivalent perspectives of the ``trace condition," momentum-space holomorphicity, and real-space ``vortexability." We will review the utility of each of these perspectives. Finally, we will conclude by developing certain technical tools for working with ideal bands that we will use in the subsequent decomposition section.

\begin{figure}
    \centering
    \includegraphics[width=0.9\linewidth]{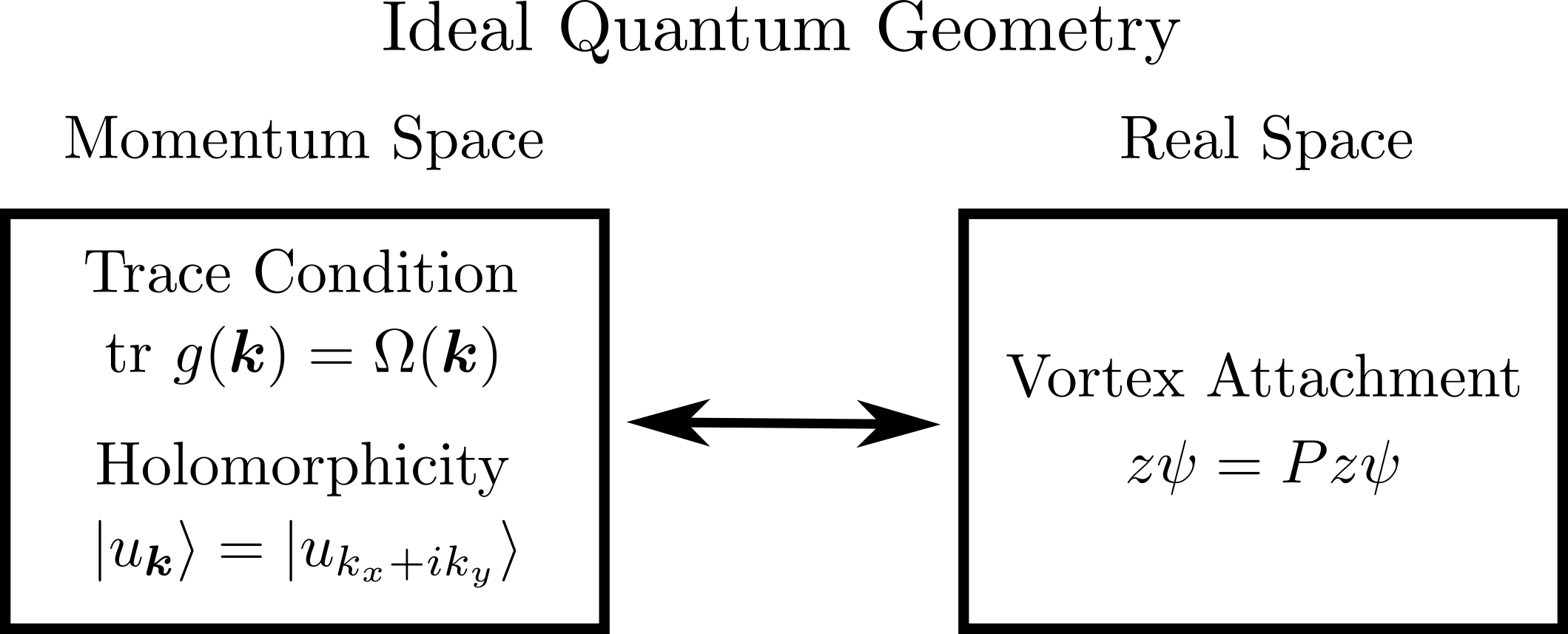}
    \caption{Equivalent descriptions of ideal quantum geometry in momentum space and real space corresponding to conditions (i)-(iii) in the main text. In momentum space we have (i) trace condition, useful for characterizing violations of ideal band geometry, and (ii) momentum holomorphicity, useful for analytic understanding of the single particle wavefunctions. In real space we have (iii) vortex attachment which enables us to directly construct many-body FCI states \eqref{eq:manybody_vortexattach} that are the ground state in short-range interaction potentials. }
    \label{fig:flow}
\end{figure}

Consider a set of bands with total Chern number $C \geq 0$ and periodic wavefunctions $u_{\bk a}(\br) = e^{-i \bk \cdot \br} \psi_{\bk a}(\br)$, where $a$ is a band index. The set of bands is \emph{ideal} if they satisfy the trace condition defined as one of the following three equivalent conditions~\cite{vortexability}:

\begin{enumerate}[(i)]
    \item The inequality $\tr g(\bk) \geq \Omega(\bk)$ is saturated, where $g_{\mu \nu}(\bk) = \Re \eta_{\mu \nu}(\bk)$ is the Fubini Study metric and $\Omega(\bk) = -\epsilon^{\mu \nu}\Im \eta_{\mu \nu}$ is the Berry curvature. \cite{parameswaranFractionalQuantumHall2013,jacksonGeometricStabilityTopological2015,meraKahlerGeometryChern2021,meraEngineeringGeometricallyFlat2021}. Here, $\eta$ is the quantum metric defined by
    \begin{equation}
        \eta_{\mu \nu}(\bk) = \sum_a \bra{ \partial_{k_\nu} u_{\bk a} } Q(\boldsymbol{k}) \ket{\partial_{k_\mu} u_{\bk a}}
    \end{equation}
    and $Q(\boldsymbol{k}) = I - \sum_{\bk a} \ket{u_{\bk a}} \bra{u_{\bk a}}$.
    
    \item The periodic wavefunctions $\ket{u_{\bk a}} = e^{-i\bk\cdot \br} \ket{\psi_{\bk a}}$ may be chosen to be holomorphic functions of $k = k_x + i k_y$ in a suitable choice of gauge. Non-unitary gauge transformations are generically required to reach this gauge, and the wavefunctions $\ket{u_{ka}}$ are generically not orthonormal.
    \item The Bloch wavefunctions $\psi$ satisfy $z \psi = \P z \psi$ where $z = x+iy$ and ${\cal P} = \sum_{\bk a} \ket{\psi_{\bk a}}\bra{\psi_{\bk a}}$ is the projector onto the bands of interest, written here using an orthonormal basis $\ket{\psi_{\bk a}}$.
\end{enumerate}
See Ref.~\cite{meraKahlerGeometryChern2021} for the equivalence of (i) and (ii) and detailed description of the associated K\"{a}hler geometry on the Brillouin Zone (BZ). 
The condition (iii) and its generalizations, together with its relationship to conditions (i) and (ii), is discussed in Ref.~\cite{vortexability}.
Pictorially, these relations are shown in \figref{fig:flow}. See also Ref. \cite{simonContrastingLatticeGeometry2020} which argued for a generalized version of condition (i); we will only use the traditional version here.

We note that the  $z \psi = \P z \psi$ originates directly from the property \eqref{eq:ideal_band_origin} of chiral graphene multilayers \cite{vortexability}. If $\psi$ is in the band of interest, the zero mode space of $\cal{D}$, so is $z \psi$ by \eqref{eq:ideal_band_origin}; we calculate ${\cal D}(z \psi) = z {\cal D} \psi = 0$. Condition (ii) may also be understood from the fact that ${\cal D}(\partial_x, \partial_y) = {\cal D}(-2i\ov{\partial})$~\cite{ledwithFractionalChernInsulator2020a,ledwithFamilyIdealChern2022}. On the periodic wavefunctions $u_\bk$ the relevant zero mode operator is $e^{-i \bk \cdot \br}{\cal D}(-2i\ov{\partial}) e^{i \bk \cdot \br} = {\cal D}(-2i \ov{\partial} + k_x + i k_y)$. Because the zero mode operator defining the periodic wavefunctions $u_{\bk}$ only depends on $k_x + i k_y$, we may choose the wavefunctions $u_\bk$ to only depend on $k_x + i k_y$. We therefore conclude that the flat bands of magic angle chiral multilayers host ideal quantum geometry.

 Each of the above statements would be useful for a different purpose. Statement (i) provides a way to quantify the violation of ideality as $\tr(g) - \Omega$~\cite{jacksonGeometricStabilityTopological2015, ledwithFractionalChernInsulator2020a, parkerFieldtunedZerofieldFractional2021}. Statement (iii) was recently interpreted~\cite{vortexability,ledwithFamilyIdealChern2022} as a ``vortexability" condition: we may attach a vortex $z$ to a wavefunction while remaining within the band of interest: $z\psi = \P z\psi$. 
 
It was recently shown \cite{ledwithFamilyIdealChern2022,vortexability} that the vortex attachment $\psi \to z \psi = \P z\psi$ enables us to generalize the Laughlin state construction to any ideal band with any Chern number $C$\,$\neq$\,$0$. For example, starting with a Slater determinant state $\ket{\Psi_{\text{Slater}}}$ we may construct
\begin{equation}
    \ket{\Psi_{2s}} = \prod_{i<j} (z_i - z_j)^{2s} \ket{\Psi_{\text{Slater}}}
    \label{eq:manybody_vortexattach}
\end{equation}
for some $s$. We note that \eqref{eq:manybody_vortexattach} should be understood on the infinite planar or expandable disk geometry, such that the electron density is allowed to change with the inclusion of the Jastrow factor $\prod_{i<j} (z_i - z_j)^{2s}$. While ordinarily the Jastrow factor would introduce weight in remote bands, leading to increased kinetic energy, the condition (iii) ensures that this does not happen in ideal bands.
The probability density $\abs{\Psi_{2s}}^2$ has a zero of order $\geq 4s$ when two particles come together which minimizes the interaction energy and ensures that \eqref{eq:manybody_vortexattach} is a ground state in the limit of short range repulsive interactions \cite{trugmanExactResultsFractional1985a,vortexability,wangExactLandauLevel2021b}.
While this argument makes it clear that ideal FCI states exist in all ideal bands, a more detailed understanding of the wavefunctions is necessary to understand the character of the FCI states \eqref{eq:manybody_vortexattach}.

In this work we use statement (ii) to characterize the single particle wavefunctions and ensuing many-body states of single ideal bands with $C$\,$>$\,$1$ which, among other results, enables us to directly characterize the FCIs generated by \eqref{eq:manybody_vortexattach}. 

We consider a single ideal band with primitive lattice vectors as $\ba_1$ and $\ba_2$ and the corresponding reciprocal lattice vectors as $\bG_1$ and $\bG_2$ where $\bG_i \cdot \ba_j = 2\pi \delta_{ij}$.
We shall always work with holomorphic, non-normalized,  wavefunctions $u_\bk = u_k = u_{k_x + i k_y}$. We furthermore ask that these wavefunctions are smooth throughout the BZ~\footnote{We may always do this because we can remove a pole at $k_0$ through gauge transformations $u_k \to f(k-k_0) u_k$ where $f$ is some quasi-periodic holomorphic function with $f(0)=0$ (e.g. Jacobi theta functions of the first kind).}. The non-trivial topology of the band is encoded by the boundary conditions 
\begin{equation} \label{eq:BDC}
    \psi_{\bk+\bG}=\Xi_\bG(k)\psi_{\bk},  \qquad \Xi_\bG(k)=\frac{u_{k+G}}{u_{k}}e^{i\bG\cdot \br}
\end{equation}
where the second equation guarantees that $\Xi(k)$ is holomorphic in $k$.

The Chern number of a single band is given by an integral of the Berry connection around the boundary of the BZ~\cite{wangExactLandauLevel2021b,ledwithStrongCouplingTheory2021,ledwithFamilyIdealChern2022}:
\begin{equation}
\begin{aligned}
    2\pi i C & =\log \Xi_{\bG_1}(k+G_2) - \log \Xi_{\bG_1}(k) \\
    & + \log \Xi_{\bG_2}(k) - \log \Xi_{\bG_2}(k+G_1) .
    \end{aligned}
    \label{Chern_from_BCs}
\end{equation}
For a multiband system $\Xi_\bG(k)$ is a matrix and the replacement $\Xi \to \det \Xi$ should be made in \eqnref{Chern_from_BCs} for the total Chern number \cite{vortexability}. Importantly, this can be interpreted as the phase winding of the wavefunction around the BZ.

For a single band with Chern number $C>0$ it is always possible to choose a gauge (see SI) such that 
\begin{equation}\label{gauge}
\begin{aligned}
    \Xi_{\bG_1}(k)&=\xi_{C,k_0}(k) \equiv \exp(2\pi i C (k-k_0)/G_2),\\ \Xi_{\bG_2}(k)&=1    
\end{aligned}
\end{equation}
for some $k_0$ in the reduced first BZ: $k_0 \equiv k_0 + \bG_i/C$.  We call boundary conditions with this form \textit{canonical}. The choice \eqref{gauge} leads to Chern number $C$ in \eqref{Chern_from_BCs}. We will make this choice for our ideal band of interest. The generality of \eqref{gauge} follows from the classification of holomorphic line bundles on the torus, as we explain in the SI.

The boundary conditions \eqref{gauge} are remarkably restrictive. Indeed, by fixing a gauge choice similar to  \eqref{gauge}, it can be shown that single particle wavefunctions for an ideal $C=1$ band take the following form~\cite{wangExactLandauLevel2021b} (see SI for a review of this argument)
\begin{equation}
    \psi_{\bk \alpha}(\br) = \psi_{\bk - \delta\bk_0}^{(\LLL)}(\br) \N_\alpha(\br),
    \label{eq:ChernOneClassification}
\end{equation}
where $\delta \bk_0 = \bk_0 - \bk_0^\LLL$.
The equality \eqref{eq:ChernOneClassification} readily yields an interpretation of the FCIs \eqref{eq:manybody_vortexattach} as generalized Laughlin states, modified from the usual Laughlin state through a periodic density modulation generated by $\sqrt{\sum_{\alpha} \abs{\N_\alpha(\br)}^2}$.

To understand the FCIs associated with higher Chern bands, we need to understand the wavefunctions in a manner analogous to \eqref{eq:ChernOneClassification}. However, a key aspect of the classification \eqref{eq:ChernOneClassification} is the separation of the $\bk$ dependence and the non-positional orbital $\alpha$ dependence. For higher Chern bands these are in general entangled and the resulting wavefunctions are much harder to work with as a result.

Instead, we will opt to \emph{decompose} higher Chern bands into $C=1$ bands of the form \eqref{eq:ChernOneClassification}.\par
\section{Decomposition of an ideal Chern number $2$ band}\label{decomp}
In this section, we provide a framework of decomposing an ideal higher Chern band into athe basis of ideal Chern $1$ bands where interesting interaction-driven physics becomes manifest.  
In particular, we illustrate a non-unitary $\SU(C)$ action on the wavefunctions, which will be promoted to an ``emergent'' symmetry on the ground state manifold of Hamiltonians with short-ranged interactions at particular fillings (see Sec.~\ref{sec:cdw} and \ref{sec:FCIs}).
As a concrete example, we will describe the decomposition of an ideal Chern band with $C\,{=}\,2$.

\subsection{Quantum Hall Bilayer}

Before we move into the technical details of decomposing generic ideal higher Chern bands, it is useful to consider an easier example, quantum Hall multilayers, ``in reverse''.
Quantum Hall multilayers can be considered to be a system with total Chern number $C\,{>}\,1$, where each layer corresponds to a band with $C\,{=}\,1$. Below, we will show how we can combine these layer bands to form a single $C>1$ band. This decomposition will shed light on the general procedure to decompose a higher Chern band by reversing this process~\cite{wuBlochModelWave2013}.
For simplicity, we start with the simplest case ofTo illustrate further, let us focus here on quantum Hall bilayer~\cite{QHFM} with a constant magnetic field $B>0$, which has a total Chern number $2$. We define magnetic translation operators $T^{\magt}_{\bl}$ that commute with the canonical momentum and satisfy the commutation relation
\begin{equation}
    T^\magt_{\bl_1} T^\magt_{\bl_2} = e^{ i B\hat{z} \cdot \bl_1 \times \bl_2}T^\magt_{\bl_2} T^\magt_{\bl_1} 
    \label{eq:magnetic_transl_commutator_main}
\end{equation}
If we use a unit cell spanned by lattice vectors $\bR_1$ and $\bR_2$ that encloses $2\pi$ magnetic flux such that the magnetic translations $T^\magt_{\bR_1}$ and $T^\magt_{\bR_2}$ commute, the system can be described as two $C=1$ bands with explicit $\SU(2)$ action of Pauli matrices $\tau_{x,y,z}$ acting on layer degrees of freedom. 

In order to ``stitch'' two bands into a single band of $C\,{=}\,2$, let us instead consider magnetic translations $T^{\text{mag}}_{\ba_1}$ and $T^{\text{mag}}_{\ba_2}$ that enclose $\pi$ flux, $B\norm{\ba_1 \times \ba_2}$\,=\,$ \pi$~\cite{wuBlochModelWave2013}. These translations anticommute, but if we combine them with layer operators
\begin{equation}
    T_{\ba_1} = \tau_x T^\magt_{\ba_1},\,\, 
    T_{\ba_2} = \tau_z T^\magt_{\ba_2}, \,\,
    T_{\ba_i} \psi_\bk(\br) = e^{i \bk \cdot \br } \psi_\bk(\br)
    \label{eq:Ceq2basis}
\end{equation}
then the translations $T_{\ba_i}$ commute and we may define Bloch states as above. Since the translations enclose half the usual area, the two $C$\,=\,$1$ LLLs on each layer combine to form a single $C$\,=\,$2$ band with wavefunctions $\psi_{\bk}(\br)$.
In this language, $\nu $\,=\,$ 1$ quantum Hall ferromagnets of the bilayer system become \emph{charge density waves} in the newly defined unit cell since they break at least one of the translation symmetries $T_{\ba_i}$. For example, polarizing the top layer, $\expval{\tau_z} \neq 0$, corresponds to breaking the translation symmetry $T_{\ba_1} \propto \tau_x$ since it anticommutes with $\tau_z$. 

We remark that the layer operators $\tau_i$ can be expressed in terms of lattice translations $T_{\ba_i}$ and magnetic translations:
\begin{equation}
    \tau_z = (T^\magt_{\ba_2})^{-1}T_{\ba_2}, \,\, \tau_x =  (T^\magt_{\ba_1})^{-1}T_{\ba_1},\,\,\tau_y =  (T^\magt_{\ba_3})^{-1}T_{\ba_3}
    \label{eq:recover_SU2}
\end{equation}
where $\ba_3 \equiv -(\ba_1 + \ba_2)$. This is the relation we want to use to construct $\SU(2)$ flavor actions when we decompose ideal higher Chern bands with $C=2$. However, there is a problem: we do not have magnetic translation symmetries defined for an arbitrary higher Chern band. Therefore, in order to use the above relation to construct flavor actions, we have to define the magnetic translations in an arbitrary setting first.
To do so, we first notice that the commutation relation in \eqref{eq:magnetic_transl_commutator_main} implies that $T^\magt_\bl$ shifts the Bloch momenta associated to $T_\ba$. 
We may thus identify magnetic translations as momentum translations:

\begin{equation}
    T^\magt_\bl \psi^\LLL_\bk(\br)  = c^\LLL_\bl(\bk) \psi^\LLL_{\bk - B\hat{\bz} \times \bl}(\br).
    \label{eq:magneto_momentum_transl}
\end{equation}
Here $\psi^{\LLL}_\bk$ are $C$\,=\,$1$ Bloch states of a single LLL with reciprocal lattice vectors $\bQ_{1,2}$ corresponding to a $2\pi$ flux lattice: $B^{-1}\norm{\bQ_1 \times \bQ_2} = 2\pi = 2 B\abs{\ba_1 \times \ba_2}$. 
The coefficient $c^\LLL_\bl(\bk)$ is not unique, depending on the momentum space gauge: under $\psi_\bk \to \lambda_\bk \psi_\bk$ we have $c^\LLL_\bl(\bk) \to \lambda_\bk \lambda_{\bk  - B\hat{\bz} \times \bl}^{-1} c^\LLL_\bl(\bk)$. 

However, we can single out a unique decomposition and choice of $\SU(2)$ action through the use of a gauge-fixed $c_\bl(\bk)$ due to our ability to pick a holomorphic gauge \eqref{gauge}.
Indeed, in the gauge \eqref{gauge}, where the LLL has $\bk_0^\LLL = -(\bQ_1 + \bQ_2)/2$, the magnetic translation operators of the lowest Landau level have the action \eqref{eq:magneto_momentum_transl} with (see SI)
\begin{equation}
\begin{aligned}
    c^\LLL_\bl(\bk) & = e^{i \bk_0^\LLL \cdot \bl} c^\naut_\bl(\bk - \bk_0^\LLL),\\
    c^\naut_\bl(\bk) & = 
    e^{i \bk \cdot \bl} e^{\frac{i}{2}(\bQ_1 + \bQ_2) \cdot \bl}e^{\frac{\pi B l}{Q_2} - \frac{i}{2} \left(\frac{\ov{Q}_2}{Q_2}l + \ov{l} \right)\left(k - \frac{iBl}{2}\right) } 
    \end{aligned}
    \label{eq:magneto_momentum_transl_gauge}
\end{equation}
 We will soon use \eqref{eq:magneto_momentum_transl}, with $c_\bl(\bk)$ given in \eqref{eq:magneto_momentum_transl_gauge}, to recover $\SU(2)$ operators \eqref{eq:recover_SU2} in an arbitrary ideal $C=2$ band. Note that we have written \eqref{eq:magneto_momentum_transl_gauge} in a manner that will enable a straightforward generalization to bands with $\bk_0 \neq \bk_0^\LLL$.
We will need the identities
\begin{equation}
\label{eq:cnaut_identities}
\begin{aligned}
   e^{i \bk_0 \cdot \bR_2}c^\naut_{\bR_2}(\bk - \bk_0) & = -e^{i \bk \cdot \bR_{2}}  \xi^{-1}_{1,k_0}(k), \\
   e^{i \bk_0 \cdot \bR_1/C}c^\naut_{\bR_1/C}(\bk - \bk_0) & = e^{i \bk \cdot \bR_1/C}
   \end{aligned}
\end{equation}
where $\bR_i \cdot \bQ_j = 2\pi \delta_{ij}$ are the associated lattice vectors and $C, \bk_0$ are arbitrary. In deriving the first equation we used $\hat{\bz} \cdot \bQ_1 \times \bQ_2 = 2\pi B$. Note that $\xi_{1 k_0}(k) = e^{2\pi i C(k-k_0)/Q_2}$ is the $k$-space boundary condition defined in \eqref{gauge}.

\begin{figure*}[!t]
  \begin{center}
  \includegraphics[width=0.7\textwidth]{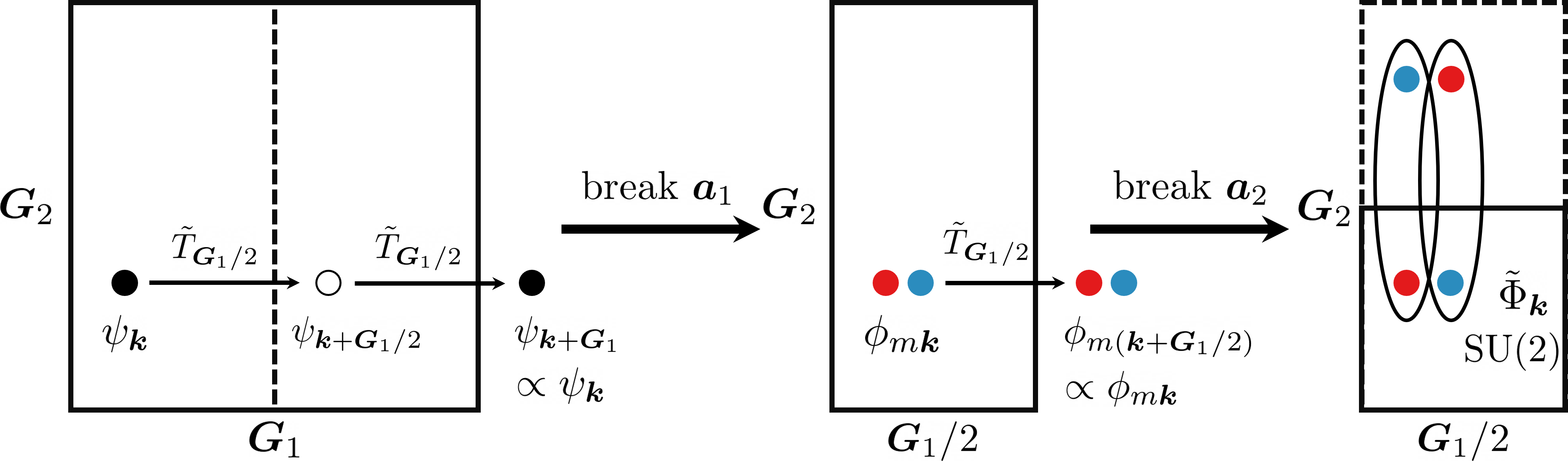}
  \end{center}
  \caption{Schematic figure that shows the decomposition process in two steps. The first step corresponds to the diagonalization of boundary conditions from the original $C$\,=\,$2$ band (shown in black and white) to form two ideal $C$\,=\,$1$ bands $\phi_{m\bk}$ (shown in red and blue), which breaks translation symmetry in $\ba_1$. The second step identifies wavefunctions at different momenta as an $\SU(2)$ multiplet $\tilde{\phi}_{m\bk}$; at each bloch momentum in quartered Brillouin Zone, there are two $\SU(2)$ multiplets that together give rise to a four component spinor $\tilde{\Phi}_\bk$ \eqref{eq:tildephi}.}
  \label{fig:decomp}
\end{figure*}

\subsection{Decomposing a general ideal Chern band}

With the understanding of the quantum Hall bilayer, we now decompose an arbitrary ideal Chern band with $C$\,=\,$2$ into two ideal $C=1$ bands that admit an $\SU(2)$ action. Note that, for a holomorphic Chern band of interest we do not generically have magnetic translation symmetry $T^\magt_\bl$ defined at the UV scale unlike for the case of LLLs in quantum Hall problems.
Nonetheless we will use \eqref{eq:recover_SU2} to recover an $\SU(2)$ action in any ideal $C$\,=\,$2$ band through a more generalizable form of $T^\magt_\ba$.

Consider an ideal $C$\,=\,$2$ band defined on a lattice with lattice vectors $\ba_{1,2}$ and reciprocal lattice vectors $\bG_{1,2}$. As a naive attempt, one can decompose the band into two $C$\,=\,$1$ bands by doubling the unit cell, for example, in the $\ba_1$ direction; the new lattice vectors are $2\ba_1$ and $\ba_2$ and the corresponding reciprocal lattice vectors are $\bG_1/2$ and $\bG_2$. Accordingly, a single-particle state in the new BZ is expressed by a doublet $(\psi_\bk, \psi_{\bk+\bG_1/2})^T$, and the boundary condition \eqref{eq:BDC} takes the following form
\begin{equation}
    \Xi_{\bG_1/2}(k)=\left(\begin{array}{cc}
        0 & 1 \\
        \xi_{2,k_0}(k) & 0
    \end{array}\right),\quad \Xi_{\bG_2}(k)=\left(\begin{array}{cc}
        1 & 0 \\
        0 & 1
    \end{array}\right)
\end{equation}  
However, note that the $\bG_1/2$ boundary condition is non-diagonal, implying that two bands are not really independent; for example, they may not be separately filled. 
In order to fully decompose the band, we seek holomorphic combinations of $\psi_\bk, \psi_{\bk+\bG_1/2}$ that \emph{diagonalize} the boundary conditions; furthermore, we require that the new, hybridized, bands have the canonical boundary conditions \eqnref{gauge}.
To achieve this, we define a matrix $U_k$ such that the transformed wavefunctions
\begin{equation}
    \left(\begin{array}{cc}
         \phi_{0\bk}\\
         \phi_{1\bk} 
    \end{array}\right)=U_k    \left(\begin{array}{cc}
         \psi_{\bk}\\
         \psi_{\bk+\bG_1/2} 
    \end{array}\right)
\end{equation}
satisfy our requirements. In doing so, we require $U_k$ to be holomorphic so that $\phi$s respect the trace condition. To ensure the diagonal boundary condition are canonical for $\phi$ bands, $U_k$ satisfies
\begin{equation}
\begin{aligned}
    U_{k+G_1/2}\Xi_{\bG_1/2}(k)U_k^{-1}&=\left(\begin{array}{cc}
        \xi_{1,g_0}(k) & 0 \\
        0 & \xi_{1,g_1}(k)
    \end{array}\right),\\ U_{k+G_2}\Xi_{\bG_2}(k)U_k^{-1}&=\left(\begin{array}{cc}
        1 & 0 \\
        0 & 1
    \end{array}\right)
\end{aligned}
\end{equation}
We need to solve for $g_0$ and $g_1$.

We find that there is a unique choice of $U_k$(See \ref{Uk}): 
\begin{equation}
    U_k=\left(\begin{array}{cc}
        1 & \xi_{1,g_0}^{-1}(k) \\
        1 & \xi_{1,g_1}^{-1}(k)
    \end{array}\right)
\end{equation}
which yields $g_0=k_0+G_1/4$, $g_1=g_0-G_2/2$. Thus the boundary conditions for ${\phi}_{m\bk}$ will be: \begin{equation}\label{solution}
\phi_{m(\bk+\bG_1/2)}=\xi_{1,g_0-mG_2/2}(k)\phi_{m\bk},\,\,\phi_{m(\bk+\bG_2)}=\phi_{m\bk}
\end{equation}Note that the $\phi_m$ bands are exchanged under $T_{\ba_1}$: 
\begin{equation}
    T_{\ba_1} \phi_{m \bk} = e^{i \bk \cdot \ba_1} \phi_{(m+1) \bk}
\end{equation}
where $m \equiv m+2$ is understood modulo two.

There is a subtlety in this ``full'' decomposition. The $\phi$ bands in general do not form an orthogonal basis. Orthogonality would only be obvious if $\psi_\bk$ and $\psi_{\bk + \bG_1/2}$ had equal norm (which they do not since we have chosen a holomorphic basis) and if the matrix $U_k$ was unitary. In principle these two sources of non-orthogonality could cancel out, but Ref.~\cite{ledwithFamilyIdealChern2022} showed that this is true if and only if the Berry curvature satisfies $\Omega(\bk) = \Omega(\bk + \bG_1/2)$. However, this does not hold in general and is specifically untrue for the particular class of chiral graphene models we are considering. Nonetheless, we will see that this non-orthogonality does not hinder us from understanding interaction-driven physics based on decomposed basis.

Now that we have constructed two ideal bands, we want to hybridize them to construct an entire manifold of ideal bands. Soon we will see that the hybridization grants us the ability to access translation breaking states in all directions. Importantly, we want to hybridize two bands with the same boundary conditions in order to keep the boundary condition diagonal.
Note that while $\phi_{0\bk}$ and $\phi_{1 \bk}$ have different boundary conditions, $\phi_{0(\bk + \bG/2)}$ and $\phi_{1\bk}$ have the same boundary conditions (see \eqref{solution}).

We therefore define
\begin{equation}\label{eq:momentumshift}
    \tilde{\phi}_{0\bk}\equiv\phi_{0\bk + \bG_2/2},\quad \tilde{\phi}_{1\bk}\equiv\phi_{1\bk}.
\end{equation}

Since we are now hybridizing states with momenta shifted by $\bG_2/2$ we should fold our Brillouin Zone accordingly. We therefore obtain \emph{four} bands which may be written as
\begin{equation}
    \tilde{\Phi}_\bk\equiv\Big( \tilde\phi_{0\bk},
         \tilde\phi_{1\bk},
         \tilde\phi_{0\bk+\bG_2/2},
         \tilde\phi_{1\bk+\bG_2/2} \Big)^T
         \label{eq:tildephi}
\end{equation}
which comprises of two pseudospin doublets from momentum points related by $\bG_2/2$. While we now have four bands, corresponding to the fact that we have folded the Brillouin Zone in half twice, these should be understood as two groups of two bands each -- for example the first and third band taken together resemble the LLL with a $4\pi$ magnetic flux in our now-quadrupled unit cell spanned by $2\ba_{i}$. These subbands cannot be filled independently without further hybridization; they are related by a shift by a reciprocal lattice vector as $\tilde{\Phi}_{1, \bk + \bG_2/2} = \tilde{\Phi}_{3, \bk}$ and therefore cannot be individually defined on the first BZ. We will  often work with $\tilde{\phi}_{m \bk}$ as if they were two bands in a Brillouin Zone spanned by $\bG_1/2$ and $\bG_2$, but it is important to emphasize that this implicitly involves a momentum shift \eqref{eq:momentumshift} and states built out of $\tilde{\phi}_{m \bk}$ can break $T_{\ba_2}$.

The real-space form $\tilde{\phi}_m$ is given by (see \eqref{eq:ChernOneClassification}):
\begin{equation}
    \tilde{\phi}_{m\bk}=\psi^{(\LLL)}_{\bk-\delta\bk_0'}(\br)\N^{(m)}(\br).
    \label{eq:classify_decomposed_bands}
\end{equation}
where $\bk_0'=\bg_0+\bG_2/2$ and $\delta \bk_0' = \bk_0' - \bk_0^\LLL$. Note that the momentum shift from $\phi$ to $\tilde{\phi}$, which made the boundary conditions $m$-independent, ensured that the origin of momentum space in the LLL part of $\tilde{\phi}$ is $m$-independent. On the other hand, the momentum shift implies $T_{\ba_2} \tilde{\phi}_{m\bk} = e^{i \bk \cdot \ba_2} (-1)^{m+1} \tilde{\phi}_{m \bk}$, such that $\N^{(m)}(\br)$ differ by a sign $(-)^m$ under translations by $\ba_2$. We will see that using $\tilde{\phi}_{m\bk}$ we can construct states that break $T_{\ba_1}$, $T_{\ba_2}$, or both. The actions of translation operators $T_{\ba_i}$ on $\tilde{\Phi}_\bk$ are
\begin{equation}\label{eq:Translation}
\begin{aligned}
    T_{\ba_2}&=-e^{i\bk\cdot \ba_2}  \mu_z \otimes \tau_z,\\
    T_{\ba_1}&=e^{i\bk\cdot \ba_1}  \mu_x \otimes  \tau_x\\
    T_{\ba_3}&=e^{i\bk\cdot \ba_3}  \mu_y \otimes  \tau_y
\end{aligned}
\end{equation}
where $\tau$ matrices act on the pseudospin index, the first and third versus the second and fourth components of $\tilde{\Phi}$, and $\mu$ matrices act on the $\bG_2$ folding index, the first two components versus the second two components.

Inspired by the quantum Hall bilayer, we will combine the magnetic translations along with real space translations to derive an $\SU(2)$ action. However, since there is no physical magnetic field, there is no immediate notion of magnetic translation. Nevertheless, we can borrow the definition of the magnetic translation as a momentum shift (cf.\eqref{eq:magneto_momentum_transl},\eqref{eq:magneto_momentum_transl_gauge}): 
\begin{equation}
    T^\magt_\bl \tilde{\phi}_{m \bk}(\br)  \equiv e^{i \bk_0' \cdot \bl} c^\naut_{\bl}(\bk- \bk_0') \tilde{\phi}_{m \bk - B_{\text{eff}} \hat{\bz} \times \bl}(\br) 
    \label{eq:magnetic_transl_phi_t}
\end{equation}
where  $B_{\text{eff}}$ is such that $B_{\text{eff}} \norm{2\ba_1 \times \ba_2} = 2\pi$ and $c_\bl(\bk)$ is given in \eqref{eq:magneto_momentum_transl_gauge}. Note that \eqref{eq:magnetic_transl_phi_t} may be interpreted as a magnetic translation acting on the just the LLL part of $\tilde{\phi}$ in \eqref{eq:classify_decomposed_bands}. Note that $T^\magt_\bl$ is generically neither unitary nor a symmetry.

In accordance with \eqref{eq:recover_SU2} we define
\begin{equation} \label{eq:SU2action}
    \mathsf{s}_z  = (T^\magt_{\ba_2})^{-1} T_{\ba_2},  \mathsf{s}_x  = (T^\magt_{\ba_1})^{-1}T_{\ba_1},  \mathsf{s}_y  = (T^\magt_{\ba_3})^{-1}T_{\ba_3}
\end{equation}
with $\ba_3=-(\ba_1 + \ba_2)$ (see e.g. Fig. \ref{fig:ferrosphere}, though the above discussion does not require $C_3$ symmetry). The magnetic translations then correspond to translations by half reciprocal lattice vectors; we have
$-B_{\mathrm{eff}}\hat{\bz} \times \ba_2 = \bQ_1 = \bG_1/2$ and $-B_{\mathrm{eff}}\hat{\bz} \times \ba_1 = - \bQ_2/2 = -\bG_2/2$.
Accordingly, the action of \eqnref{eq:SU2action} on the spinor $\tilde{\Phi}$ yields 
\begin{equation}
    \mathsf{s}_z = \mu_0 \otimes \tau_z, \,\,\, \mathsf{s}_x = \mu_0 \otimes \tau_x, \,\,\,\mathsf{s}_y = \mu_0 \otimes \tau_y    .
    \label{eq:action_of_s}
\end{equation}
 In deriving \eqref{eq:action_of_s} we used the action of translations on the spinor \eqref{eq:Translation}, the boundary conditions $\tilde{\phi}_{m\bk + \bG_1/2} = \xi_{1,k_0'}(k) \tilde{\phi}_{m\bk}$ and $\tilde{\phi}_{m\bk + \bG_2} = \tilde{\phi}_{m\bk}$, and the identities \eqref{eq:cnaut_identities}. 

We give a final note on translation breaking patterns of different states. Given the representation of translation \eqref{eq:Translation}, states with nonzero $\mathsf{s}_z$ polarization break $T_{\ba_1}$ and states with nonzero $\mathsf{s}_x$ polarization break $T_{\ba_2}$. Therefore, we can access all sorts of translation breaking patterns within the $\tilde{\Phi}$ basis. The procedure through which we decompose the $C$\,=\,$2$ band is shown pictorially in \figref{fig:decomp}.\par

Although in this basis the operators $\mathsf{s}_i$ look Hermitian, they are not. Indeed, the basis is written with respect to the $\tilde{\phi}_{m \bk}$ wavefunctions, and these wavefunctions are not orthogonal to each other. 
However, this does not prevent us from building interesting many-body quantum states.\par

\section{Manifold of Charge Density Waves and Skyrmion textures}\label{sec:cdw}

We now move onto the construction of many-body states. In this section, we construct a continuous manifold of topological CDW states at half filling of an ideal $C=2$ band; this manifold will be identified with a sphere of generalized quantum Hall ferromagnets. These states arise out of the $\SU(2)$ action on the two decomposed $C$\,=\,$1$ bands. We will see that the $\SU(2)$ action on the band becomes an emergent symmetry of the ground state manifold. This manifold admits skyrmion textures of CDWs that carry electric charge.

We consider a repulsive density-density interaction in the short-range interaction limit. For this section, this amounts to choosing a contact interaction
\begin{equation}
    H = \frac{V_0}{2}\sum_{i<j} \delta(\hat{\br}_i - \hat{\br}_j)
    \label{eq:contact_hamiltonian}
\end{equation}
with $V_0>0$.
We have written the interaction in first quantization and $\hat{\br}_i$ is the position operator associated to particle $i$.

We now construct the manifold of charge density waves that emerges as the zero-energy ground state manifold of \eqref{eq:contact_hamiltonian}. Consider fully filling a single band formed by a generic linear combination $\sum_{m} \alpha_m \tilde{\phi}_{m, \bk}$ of the $\tilde{\phi}_{m}$ bands constructed in the previous section. The band formed by the linear combination will have wavefunctions
\begin{equation}
\begin{aligned}
    \tilde\varphi_{\theta,\phi,\bk}(\br)&\equiv\cos(\theta/2)\tilde{\phi}_{0\bk}(\br)+\sin(\theta/2)e^{i\phi}\tilde{\phi}_{1\bk}(\br)\\&=\psi^{(\LLL)}_{\bk-\bk_0'}(\br)\mathcal{N}^{(\theta,\phi)}(\br)
\end{aligned}    
\label{eq:theta_phi_wfs}
\end{equation}
for some $(\theta,\phi)$ determined by $\alpha_m$ and $\mathcal{N}^{(\theta,\phi)} =\cos(\theta/2)\mathcal{N}^{(0)}+\sin(\theta/2)e^{i\phi}\mathcal{N}^{(1)} $. Thus, for every point $(\theta,\phi)$ on the Bloch sphere we obtain an ideal $C=1$ band.
Fully filling the band \eqref{eq:theta_phi_wfs}  will generate a Slater determinant state
\begin{equation}\label{eq:QHFM}
    \ket{\theta,\phi}\equiv\mathcal{A}\bigotimes_{\bk}\ket{\tilde{\varphi}_{\theta,\phi,\bk}}
\end{equation}
where $\mathcal{A}$ antisymmetrizes the particles and $\bk$ runs through the BZ with reciprocal lattice vectors $\bG_1/2$ and $\bG_2$ (the states \eqref{eq:QHFM} still generically break $T_{\ba_2}$ translation symmetry: see the discussion below \eqref{eq:tildephi}). The Slater determinant forces the momentum dependence to be antisymmetric, which, together with the form \eqref{eq:ChernOneClassification}, forces the $\br$-coordinate dependence to be antisymmetric. Thus, $\ket{\theta,\phi}$ is annihilated by the contact interaction.

We have constructed a ferromagnetic ground state $\ket{\theta, \phi}$ associated with each point $(\theta,\phi)$ on the pseudospin-sphere. Each ferromagnet is a topological CDW that is translationally symmetric with a $2\ba_1\times2\ba_2$ unit cell. At special points --- on the $X,Y,Z$ axes --- the state further preserves one of the translation symmetries $T_{\ba_i}$ for $i = 1,3,2$ (see \eqref{eq:Translation}). The order parameter sphere is graphically shown in Fig.~\ref{fig:ferrosphere}(a).

\par

\begin{figure}
    \centering
    \includegraphics[width=0.95\linewidth]{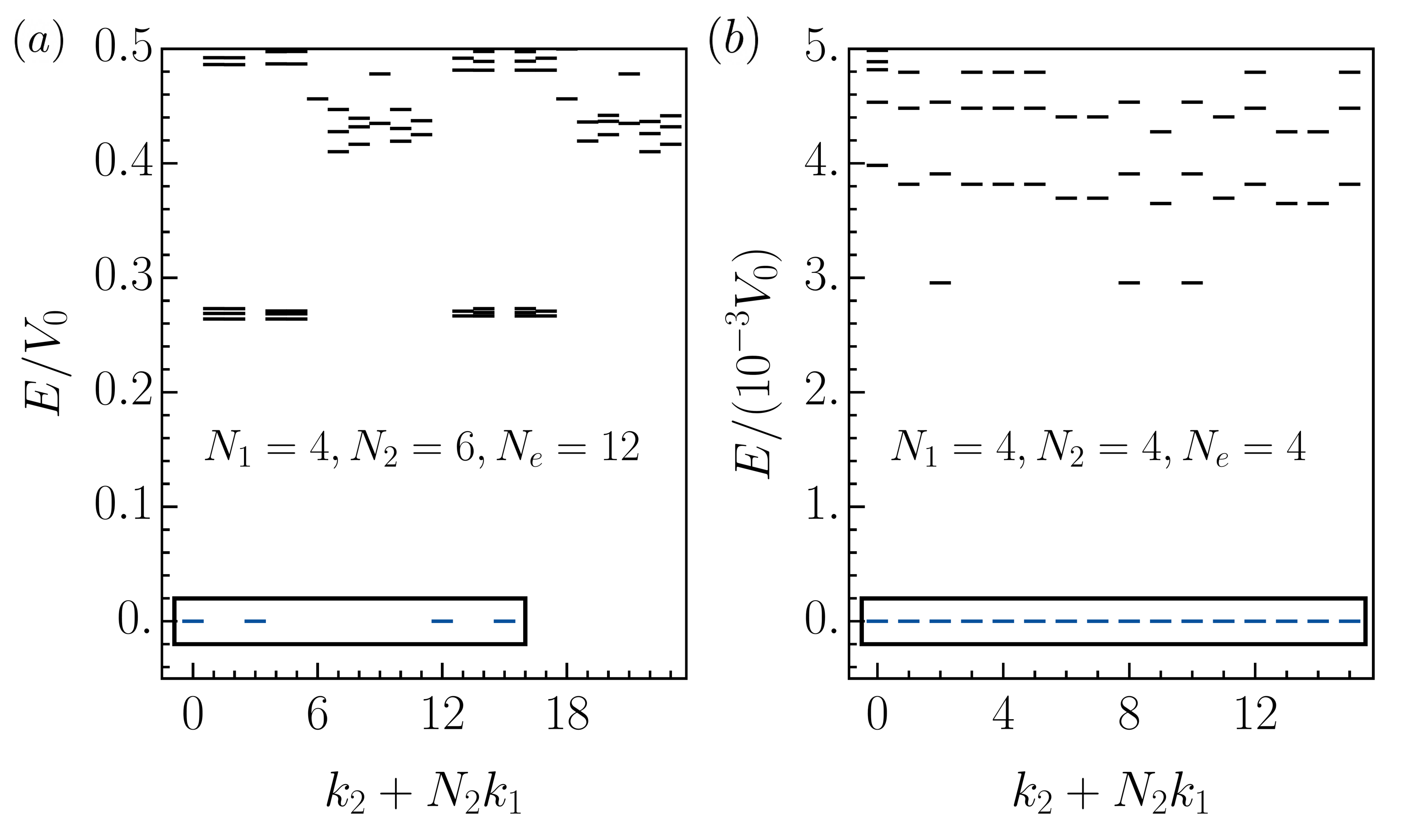}
    \caption{Momentum resolved spectrum for chiral twisted graphene multilayers with contact interaction from ED. Ground states are exact zero modes surrounded by black boxes (there are multiple zero modes within one momentum sector): they are identified with the analytically constructed charge density waves explored in this section. In panel (a): $C$\,=\,$2$, $\nu$\,=\,$1/2$, and GSD\,=\,$13$. (b): $C$\,=\,$4$, $\nu$\,=\,$1/4$ and GSD\,=\,$35$. The ground state degeneracies match exactly with Eq.~(\ref{su2}) for $C=2$ and for higher Chern numbers Eq.~(\ref{hook}).}
    \label{fig:ferro}
\end{figure}

The manifold of charge density waves appears in finite-size numerics through an extensive ground state degeneracy. The degeneracy is given by the dimension of the $\SU(2)$ irreducible representation that ferromagnetic states belong to. Such an irrep is given by the symmetric tensor product of $N_e$ spin-$1/2$ systems.
This is the spin $N_e/2$ respresentation of $\SU(2)$ whose dimension is
\begin{equation}\label{su2}
    \text{GSD}=N_e+1.
\end{equation}
In our numerical calculations, we use the ideal Chern bands of chiral twisted graphene multilayers \cite{ledwithFamilyIdealChern2022} with contact interaction.
The ground state degeneracies are confirmed via ED, see Fig.~(\ref{fig:ferro}), for systems that have an even number of $\ba_1 \times \ba_2$ unit cells in each direction.
For $m\ba_1\times n\ba_2$, where either $m$ or $n$ is odd, the $\SU(2)$ operators cannot be defined and only the special ferromagnets that preserve $T_{\ba_1}$ or $T_{\ba_2}$ for $m$ or $n$ odd respectively will survive, leading to a GSD of two. We indeed see that this is the case.\par

We now investigate the role of rotational crystalline symmetries. They further relate ferromagnets to each other, so they are identified as rotations on the order parameter sphere.
The representation of the states as points on the sphere and crystalline symmetries as rotations is summarized in \tabref{fig:ferrosphere}(b). In particular, we focus on the
three-fold rotational symmetry $C_{3z}$ relevant to graphene moir\'{e} systems~\cite{TCDW,Wilhelm2204}.
Since $C_{3z}$ permutes the lattice vectors $\ba_1 \to \ba_2 \to \ba_3$ it also permutes the states that preserve the translation symmetries $T_{\ba_{1,2,3}}$ which correspond to the $X,Z,Y$ axes of the order parameter sphere respectively. We therefore conclude that $C_{3z}$ is a rotation around the $\bn_0 = (1 1 1)^T$ axis of the order-parameter sphere. There are other rotationally symmetric axes $\bn_i$ related by translations $T_{\ba_i}$; they correspond to the rotation symmetry $C_{3z}^{(i)} = T_{\ba_i} C_{3z} T_{-\ba_i}$ that rotates about the point $(x,y) = \ba_i$. It is convenient to identify $C^{(0)}_{3z} = C_{3z}$ and $\ba_0 = 0$. 

The crystalline symmetries $C_{3z}$ and $T_{\ba_i}$ together form a subgroup of the emergent $\mathrm{SO}(3)$ symmetry of the order parameter sphere. We identify this subgroup as the alternating group on four elements $A_4$. The four elements being permuted may be taken to be the rotationally symmetric axes $\bn_i$, for $i = 0,1,2,3$. The rotations $C_{3}^{(i)}$ are the four different three-cycles, and the translations $T_{\ba_i}$ are products of two transpositions.

We now discuss topological textures of the order parameter sphere. Because $\pi_2(S^2) = \mathbb{Z}$, there are non-trivial solitonic textures of the order parameter known as skyrmions. These skyrmions carry electric charge $Ce=e$ where $C=1$ is the Chern number of each ferromagnet~\cite{girvinQuantumHallEffect1999,sondhiSkyrmionsCrossoverInteger1993}. Note that these charge $e$ skyrmions have arisen out of a $C$\,=\,$2$ band, where one may a priori expect charge $2e$ skyrmions.
We numerically plot the density profile of such a skyrmion in Fig.~\ref{fig:results}(a). We assume a skyrmion texture whose core points at $\bn_0$ and at infinity points at $-\bn_0$. The azimuthal angle and the polar angle from $\bn_0$ is assumed to be $\theta(\br)=\pi(1-\exp(-r^2/2\lambda^2)),\phi(\br)=-\phi$.

We also note that the ferromagnets still have very isotropic energy even for longer ranged interactions and realistic band geometry. This suggests skyrmion may well be the lowest energy charged excitations in this system.

\section{Adding real electron spin: $\SU(4)$ ferromagnets and skyrmion crystals}\label{sec:spin}

Inclusion of an additional degree of freedom, such as the real electron spin, greatly enriches the manifold of ferromagnets. With spin, we now have four ideal $C=1$ bands $\tilde{\phi}_{m\sigma}$, where $\sigma = \uparrow,\downarrow$ labels the physical electron spin. We obtain a zero energy ground state under a contact interaction at $\nu=1/2$ of the $\ba_1\times\ba_2$ unit cell through filling any linear combination of these four Chern bands: this space of ground states has a natural $\U(4)$ action that rotates the filled Chern band into the empty ones and can be identified as the coset manifold $\U(4)/\U(3)\times U(1)=\mathbb{CP}^3$. Here $\U(4)$ rotates the four ideal Chern bands into each other and $\U(3)$, $\U(1)$ are redundancies that correspond to changes of bases of the empty and filled bands respectively. The charge density waves of the previous section are encompassed here as states that are both pseudospin polarized and spin polarized. These form an $S^2 \times S^2$ submanifold of $\mathbb{CP}^3$.

More interesting states from the ferromagnetic manifold emerge when the pseudospin and real spin are entangled, such that both polarizations are individually zero. For example, consider filling the band described by the wavefunctions
\begin{equation}
    \tilde{\phi}_{\bk} = \frac{1}{\sqrt{2}}(\tilde{\phi}_{0, \uparrow ,\bk} - \tilde{\phi}_{1,\downarrow,\bk}).
    \label{eq:singlet_band}
\end{equation}
The state \eqref{eq:singlet_band} breaks both spin rotations as well as translations $T_{\ba_i}$. However, the combinations
\begin{equation}
    \tilde{T}_{\ba_1} = T_{\ba_1}\sigma_x, \qquad \tilde{T}_{\ba_2} = T_{\ba_{2}} \sigma_z
\end{equation}
are preserved such that the charge density of \eqref{eq:singlet_band} is symmetric in the original $1 \times 1$ unit cell. We identify these states with the tetrahedral antiferromagnets (TAFs) in~\cite{Wilhelm2204}. \par

\begin{figure}
    \centering
    \includegraphics[width=0.95\linewidth]{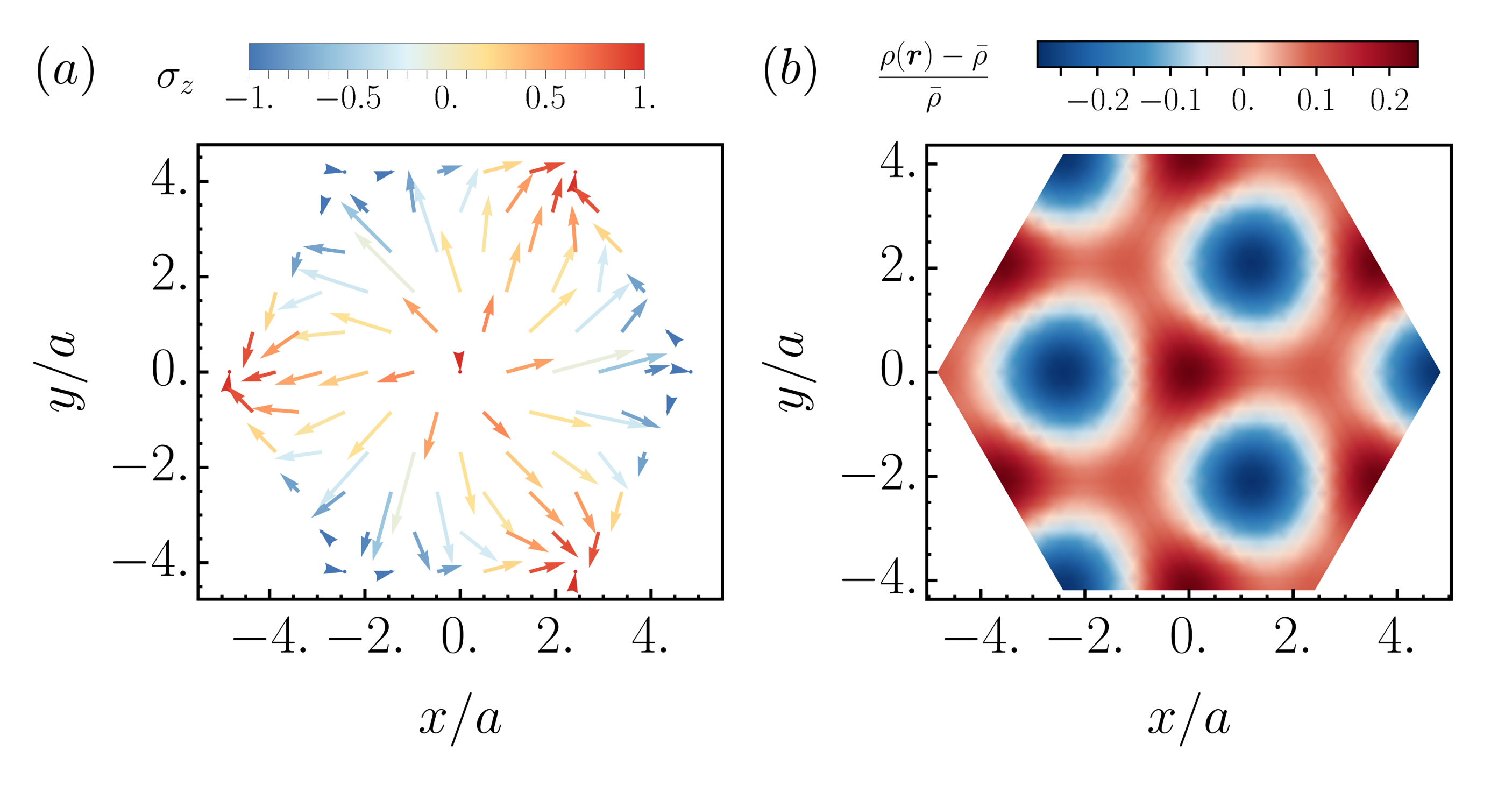}
    \caption{(a) Spin texture and (b) density profile of a TAF state: it is invariant under combinations of single translations and magnetic rotations. Thus the spin texture is not periodic in the quadrupled unit cell but the density profile is periodic. In the quadrupled unit cell, the spin texture winds around like a skyrmion, which is why this state is also identified with a skyrmion crystal. For presentation, we rotated the spin texture such that the spin is pointing in the $+z$ direction at the center of the unit cell.}
    \label{fig:TAF}
\end{figure}

We may apply spin rotations to \eqref{eq:singlet_band} to obtain all the states that have both zero spin and zero pseudospin polarization. Since there is no fixed axis in the spin texture of the TAF state, the spin rotation action is faithful. Thus, the manifold of all TAF states is equivalent to the group manifold $\mathrm{SO}(3)$. Due to its special spin texture throughout the unit cell, it can also be understood as a skyrmion crystal. In Fig. \ref{fig:TAF} we plot the charge density and spin texture of an analytically constructed skyrmion crystal state in the $2 \times 2$ unit cell. We choose a state where the spin is pointing in the $+z$ direction at the center of the unit cell for aesthetic purposes. We note that the charge density is visibly periodic with respect to the $1 \times 1$ unit cell.

\section{Fractional Chern Insulator}\label{sec:FCIs}

Having established the quantum Hall ferromagnets in the ideal $C$\,=\,$2$ band, we now investigate the various FCIs that arise in this system. We classify the FCIs into two categories: translation symmetric FCIs and translation broken FCIs. The first category corresponds to flavor singlets after the decomposition, whereas the second category corresponds to a flavor ferromagnet.

\subsection{Translation Symmetric FCIs}

Vortex attachment in ideal Chern bands ensure that FCIs exist through \eqref{eq:manybody_vortexattach}. 
However, what is the topological order and filling of the state given by~\eqref{eq:manybody_vortexattach}? At $C$\,$=$\,$1$, the expression gives a Laughlin topological order at filling $1/(2s+1)$ due to the wavefunction classification \eqref{eq:ChernOneClassification} \cite{ledwithFractionalChernInsulator2020a,wangChiralApproximationTwisted2021a}. What about bands with $C$\,$>$\,$1$?

Before discussing FCIs in higher Chern bands, it is instructive to review possible FQH states in multi-component Landau levels. For example, consider a system consisting of two LLLs, either labeled by layer or flavor index. In such a system, it is well-known that the so-called Halperin $(mmn)$ wavefunction can be written as the following~\cite{Halperin:1983zz}:
\begin{equation}
\begin{aligned}
    & \Psi^{(mmn)}_{l_1,\ldots, l_{2N}}(\br_1,\ldots,\br_{2N}) \\
    & = \mathcal{A} \bigg(\prod_{i<j = 1}^N (z_i - z_j)^m (w_i -w_j)^m (z_i - w_j)^n \\
    & \times \prod_{i = 1}^N e^{-\frac{\abs{\br_i}^2}{4\ell_B^2}} e^{-\frac{\abs{\br_{i+N}}^2}{4\ell_B^2}} \delta_{1,l_i} \delta_{2, l_{i+N}}\bigg).
    \label{eq:mmn_states}
    \end{aligned}
\end{equation}
Here we have taken the first $N$ particles to be in the first layer and the second $N$ particles to be in the second later and defined associated complex coordinates $z_i = x_i + i y_i$ and $w_i = x_{i + N} + i y_{i + N}$. The antisymmetrization operator $\mathcal{A}$ enforces antisymmetry between the coordinates $i<N$ and $i>N$; within each class antisymmetry is guaranteed by restricting $m$ to be an odd integer.
We have used the symmetric electromagnetic gauge for convenience. 
The state describes an electronic state at filling $\nu = 2/(m+n)$ with respect to a unit cell of volume $2\pi \ell_B^2$, which can be computed in the standard manner by counting the maximal power of $z$. Accordingly, $\Psi^{(110)}$ corresponds to the fully filled state with $\nu=2$.

By direct computation, one can show that the states $\Psi^{(2s+1,2s+1,2s)}$ arise from the vortex attachment \eqref{eq:manybody_vortexattach} applied to the fully filled Slater determinant state $\Psi^{(110)}$. 
Note that the states $\Psi^{(2s+1,2s+1,2s)}$ are singlets since  neither $\Psi^{(110)}$ nor the Jastrow factor in \eqref{eq:manybody_vortexattach} breaks the $\SU(2)$ flavor symmetry.

We now apply \eqref{eq:manybody_vortexattach} to a generic ideal $C$\,=\,$2$ band, where the Slater state is the fully filled state. Since an ideal $C=2$ band can be decomposed into two non-orthogonal $C=1$ bands, the natural expectation is that vortex attachment~\eqref{eq:manybody_vortexattach} yields a state analagous to $\Psi^{(2s+1,2s+1,2s)}$. We now show precisely how this works.

The fully filled state $|\Psi^{(C=2)}_{\nu = 2} \rangle$ is obtained by filling each band labeled by $m$ successively, regardless of orthogonality. Indeed under a non-unitary gauge transformation $c_\alpha \to d_\alpha = S_{\alpha \beta} c_\beta$ we have $\varepsilon^{\alpha_1 , \ldots \alpha_N} c_{\alpha_1} \ldots c_{\alpha_N} = \det S^{-1} \varepsilon^{\alpha_1 , \ldots \alpha_N} d_{\alpha_1} \ldots d_{\alpha_N}$ such that the fully filled state takes the same form in any basis regardless of orthonormality. Using the relationship~\eqref{eq:ChernOneClassification}, we obtain
\begin{equation}
\begin{aligned}
    & \Psi^{(C=2)}_{\alpha_1, \ldots \alpha_N}(\br_1, \ldots, \br_{2N}) \\
    & = \mathcal{A} \bigg( \prod_{i<j}^N (z_i - z_j) (w_i - w_j)  \\
    & \times \prod_{i = 1}^{N} e^{-\frac{\abs{\br_i}^2}{4\ell_{B_{\text{eff}}}^2}} e^{-\frac{\abs{\br_{i+N}}^2}{4\ell_{B_\text{eff}}^2}} \N^{(0)}_{ \alpha_i}(\br_i) \N^{(1)}_{\alpha_{i+N}}(\br_{i+N}) \bigg)
    \end{aligned}
    \label{eq:Ceq2_fullyfilled}
\end{equation}
where $\N^{(m)}$ are defined in \eqref{eq:classify_decomposed_bands}. As before, the antisymmetry operator $\mathcal{A}$ antisymmetrizes the particles $i<N$ with $i > N$, and we have used $z$ and $w$ to distinguish the particle positions on the top and bottom layer respectively. As in previous sections, $CB_{\text{eff}} \abs{ \ba_1 \times \ba_2} = 2\pi$ such that the original unit cell of the $C=2$ band has $\pi$ effective flux.

Then, if we act with the vortex attachment factor $\prod_{i<j} (z_i - z_j)^{2s}$, we obtain a state directly analogous to the Halperin state at the same filling
\begin{equation}
\begin{aligned}
    & \Psi^{(C=2, 2s)}_{\alpha_1, \ldots \alpha_N}(\br_1, \ldots, \br_{2N}) \\
    & = \mathcal{A} \bigg( \prod_{i<j}^N (z_i - z_j)^{2s+1} (w_i - w_j)^{2s+1} (z_i - w_j)^{2s}  \\
    & \times \prod_{i = 1}^{N} e^{-\frac{\abs{\br_i}^2}{4\ell_{B_{\text{eff}}}^2}} e^{-\frac{\abs{\br_{i+N}}^2}{4\ell_{B_\text{eff}}^2}} \N^{(0)}_{\alpha_i}(\br_i) \N^{(1)}_{\alpha_{i+N}}(\br_{i+N}) \bigg)
    \end{aligned}
    \label{eq:Ceq2_mmn}
\end{equation}
The only modification is the extra data associated with the non-positional orbital indices $\alpha$, and a periodic charge modulation generated by the $\br$-dependence of $\N^{(m)}_{\alpha}(\br)$.

The filling and topological order is independent of the functions $\N^{(m)}_{ \alpha}(\br)$: these functions simply modify the orbital and positional textures at the unit cell scale, while the filling is set by the competition between the high power of $z_i$ and the exponential decay of $e^{-\frac{\abs{\br_i}^2}{4\ell_{B_{\text{eff}}}^2}}$ at long distances. We obtain a filling of $2/(4s+1)$ electrons per $2\pi \ell_{B_{\text{eff}}}^2$ area, or $1/(4s+1)$ electrons in the $C=2$ unit cell. The states \eqref{eq:mmn_states} have a ground state degeneracy $m^2 - n^2 = 4s+1$ in the flavor singlet case. We expect this to be true for \eqref{eq:Ceq2_mmn} as well, and indeed find that it is the case in exact diagonalization. We remark that the constructed state $ \Psi^{(C=2, 2s)}$ is a flavor singlet, which implies that this family of FCI states is translation symmetric. As reported in Ref.~\cite{wangHierarchyIdealFlatbands2021}, the authors have also verified an FCI in the $C=2$ band, at filling factor $\nu=1/5$ (see Fig.~\ref{fig:results}(b)) with ground state degeneracy $5$. It is compatible with the flavor singlet Halperin $(332)$ state \eqref{eq:Ceq2_mmn} which is a translation symmetric zero-mode of the interaction $V_2\delta''(\hat{\br}_i-\hat{\br}_j)$. A straightforward extension of the above construction to $C$ decomposed flavors (see Sec. \ref{sec:higherC} and SI) yields flavor-singlet Halperin states at fillings $1/(2Cs+1)$ of the Chern $C$ band.

Note that ground state degeneracy and particle-entanglement-spectrum counting consistent with the Halperin state was obtained previously by Ref.~\cite{wangHierarchyIdealFlatbands2021}. Our analytic understanding complements this result. 

\subsection{Translation Broken FCIs}
After studying translation symmetric FCIs, we now study translation broken FCIs. Since the translation symmetric FCIs are flavor singlets, we can break translation symmetry by polarizing one of the flavors, generating ferromagnetic FCIs. These states can be obtained by applying the vortex attachment procedure \eqref{eq:manybody_vortexattach} to a quantum Hall ferromagnet \eqref{eq:QHFM}. Consider attaching $2s$ vortices to the ferromagnet:
\begin{equation}
    \ket{\Psi_{2s,\theta,\phi}} = \prod_{i<j} (z_i - z_j)^{2s} \ket{\theta,\phi}.
\end{equation}
Given the similarity of the ferromagnet to a fully filled LLL, the filling fraction would be $1/(2s+1)$ for the ferromagnet. In the original $C=2$ unit cell, the filling fraction would be $1/(4s+2)$. These states correspond to ferromagnetic Laughlin states.\par
\begin{figure}
\begin{center}
    \includegraphics[width=0.95\linewidth]{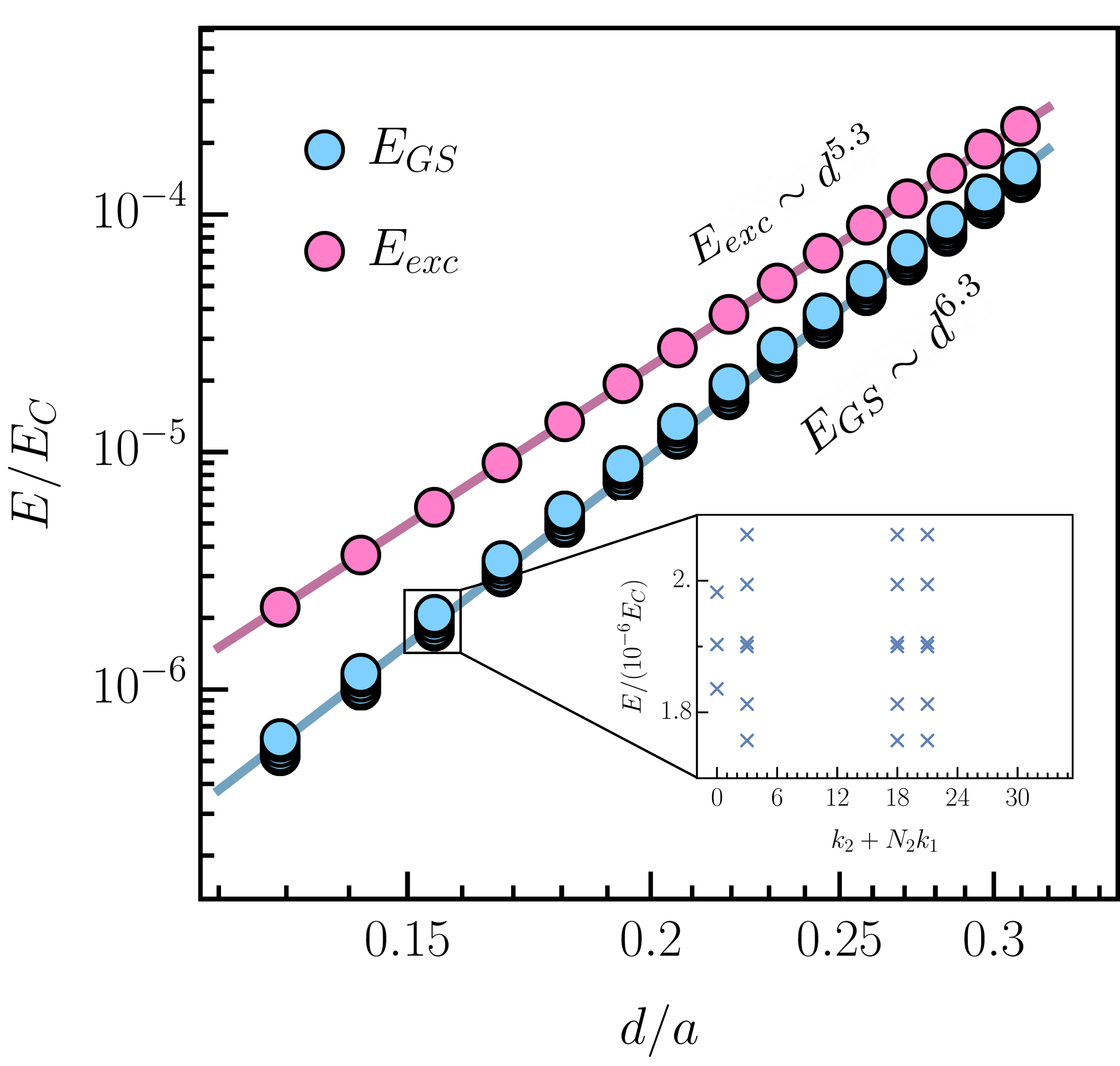}
\end{center}
    \caption{Scaling form of ground state energies $E_{\rm GS}$ and the first excited state energy $E_{\rm exc}$ for a $C=2$ band at $\nu=1/6$ with $N_1=N_2=N_e=6$ under screened Coulomb interaction at different gate distances. The dots correspond to the numerically-obtained exact diagonalization spectrum and the lines are log-linear fits to the spectrum. The extracted power laws $E\sim d^\eta$ are comparable to the theoretical expectation $\eta_{\rm GS} = 7$ and $\eta_{\rm exc} = 5$, with deviations being attributed to the finite range of $d$, in particular since very small $d$ is hard to access numerically.  
    The inset shows the spectrum of $21 = 3(N_e + 1)$ Laughlin states, which follows from the $\SU(2)$ representation theory \eqref{su2}, \eqref{eq:laughlinsu2}.}
    \label{fig:laughlin}
\end{figure}

We consider the simplest case where $s=1$, which corresponds to $\nu=1/6$. It's interesting to study whether these ferromagnetic Laughlin states will be ground states of the Coulomb interaction projected to the $C=2$ band. We consider the dual-gate screened Coulomb interaction
\begin{equation}
    V_C(\bq)= \frac{E_C}{|\bq|a}\tanh(|\bq|d)=\sum_{n\geq 0} V_n |\bq|^{2n} d^{2n+1}
\end{equation}
where $d$ is the distance from the system to the gates. Since the $V_{0,1,2}$ terms annihilate the Laughlin state, the lowest term in the Hamiltonian that contributes to energy of the Laughlin states is $|\bq|^6$. Thus we would expect a ground state energy dependence of $d^7$. On the other hand, any other state that is not the Laughlin state at this filling would have a lower order zero in the wavefunction, and in general its energy will gain contribution from the $|\bq|^4$ term, resulting in a power law of $d^5$ or lower. We perform ED to compute the power law dependence of the ground state energies on $d$ and the energy of the first excited state in Fig.~(\ref{fig:laughlin}). The numerical spectrum clearly shows that the ground states have the same power law scaling in $d$, which resembles the scaling of Laughlin states, whereas the first excited states have a different scaling behavior, which we expect to be $\sim d^5$ for sufficiently small $d$. We thus conclude that the ground states are ferromagnetic Laughlin states.\par
The numerical studies further confirm our result on the emergent $\SU(2)$ action on the ferromagnetic states. For $N_e$ electrons, there are $N_e+1$ generalized ferromagnetic states \eqref{su2}. Each ferromagnetic state will then generate $2s+1$ topologically-degenerate Laughlin states, so in total
\begin{equation}\label{eq:laughlinsu2}
    \text{GSD}\approx (2s+1)(N_e+1)
\end{equation}
which matches with the numerics in Fig. \ref{fig:results}c and Fig. \ref{fig:laughlin} for $N_e = 6$ and $s=1$. For a Chern $C>2$ band we similarly decompose into $C$ Chern $1$ bands (see Sec. \ref{sec:higherC} and SI) and make a Laughlin state in one; we obtain a filling factor $\nu=1/C(2s+1)$ and a ground state degeneracy of $(2s+1)$ multiplied by the associated ferromagnetic degeneracy \eqref{hook}.
\par

\section{Ideal $C>2$ Bands}\label{sec:higherC}
After studying the ideal $C=2$ bands in detail, we briefly summarize our results on ideal bands with $C>2$. Detailed derivations can be found in SI.\par
Ideal bands with arbitrary Chern number $C$ can also be decomposed into ideal subbands with lower Chern number by breaking translation symmetry (i.e. choosing a larger unit cell). If the unit cell is expanded by a factor of $r$, where $r$ divides $C$, then we show that there are $r$ subbands which may be hybridized such that each band is ideal and carries Chern number $C/r$. 

We will focus on the case $r=C$, where we obtain $C$ ideal Chern $1$ bands $\phi_m$ by breaking translation symmetry by $C$ times in one direction, say $\ba_1$. These bands will have different boundary conditions at the same momentum point. However, similar to the $C=2$ case, we are able to construct a hidden $\SU(C)$ action if we break translation symmetry in the other direction too, such that the final unit cell is expanded by a factor of $C$ in both directions. The bands $\tilde{\phi}_{m\bk}=\phi_{m(\bk+\bG_2/2-m\bG_2/C)}$ will again have the same boundary conditions. We may now form linear superpositions
\begin{equation}\label{ChernCmanifold}
   \phi_{v \bk}(\br)=\sum_m v_m\tilde{\phi}_{m\bk}(\br) 
\end{equation}
 where $v$ is a vector in $\mathbb{C}^C$. The states \eqref{ChernCmanifold} are generically translationally symmetric symmetric only with respect to a $C\ba_1\times C\ba_2$ unit cell. The mixed translations $(T^{\magt}_{\ba_i})^{-1}T_{\ba_i}$ generate an $\SU(C)$ action similar to the $\mathsf{s}_{x,y,z}$ operators for $C=2$ (see SI).\par
We now discuss the ground states of an ideal Chern band at $\nu=1/C$ for the contact interaction (\ref{eq:contact_hamiltonian}). We are now also able to construct a manifold of topological CDWs for this system
\begin{equation}\label{chernCferro}
    \ket{v}=\mathcal{A}\bigotimes_{\bk}\ket{\phi_{v \bk}}.
\end{equation}
Due to the redundancy $v_m \to \lambda v_m$ we identify this manifold with $\mathbb{CP}^{C-1}$.
This manifold of CDWs appear in finite size numerics too, with appropriate degeneracy counting \eqref{hook}, see  Fig.~\ref{fig:ferro}(b).\par

FCI states can also be constructed for general $C$ via the vortex attachment procedure. As with $C=2$, attaching $2s$ vortices to the fully filled Chern $C$ band constructs a flavor symmetric Halperin state at filling $\nu = 1/(2Cs+1)$.
We may also attach $2s$ vortices to a flavor polarized (translationally broken) ferromagnet
\eqref{chernCferro} to obtain a ferromagnetic Laughlin state at $\nu = 1/C(2s+1)$. 

\section{Conclusion}\label{sec:conclusion}
In this work we have explored the correlation physics in higher Chern bands, using an analytically tractable starting point, the ideal Chern bands. This approach allows us to make several rigorous statements regarding the energetics of ground states, which continue to hold even when we relax the conditions for strictly ideal bands. They provide a framework to understanding correlated states in realistic higher Chern band systems such as chiral graphene multilayers, that are already the subject of intense numerical and experimental study.

Our approach involves decomposing ideal flat Chern bands with arbitrary Chern number $C$ into $C$ ideal Chern bands with Chern number $1$. We constructed combinations of real space and momentum space translations to form a non-unitary $\SU(C)$ action between the bands. We then analytically wrote down ideal many-body ground states such as generalized quantum Hall ferromagnets, flavor-singlet Halperin states, and Laughlin ferromagnets that are exact ground states in the limit of short-range interaction potentials. We found that the $\SU(C)$ action often becomes an emergent symmetry in the ground state manifold of short range interactions.
For $C=2$, we confirm our analytic predictions with numerical evidence that these states are indeed ground states of interactions projected to the ideal flat Chern bands of chiral twisted graphene multilayers and make concrete experimental predictions for twisted graphene multilayers at various fillings. Our work opens a route to an analytic understanding of correlated phenomena in higher Chern bands.

\par
Many interesting questions arise naturally after this work. Since the decomposition in general gives non-orthogonal bands, when are the bands orthogonal? The $C=2$ case is solved in Ref.~\cite{ledwithFamilyIdealChern2022} but remains to be extended for general $C$.
Beyond ideality, how can we understand the manifold of charge density waves?
The manifold of states survives outside the ideal limit and long range interactions with suitable anisotropies, but thus far it can only be accessed through exact diagonalization\cite{Wilhelm2204}.
While we have shown FCIs exist in the ideal limit, a phase diagram with realistic parameters, including band dispersion enhancement from integrating out remote bands \cite{parkerFieldtunedZerofieldFractional2021} remains to be constructed. While there are caveats, exciting experiments have already seen $C=1$ correlated insulator phases at $\nu=3+1/2$, where realistic models predict that the ground state is in the same manifold as our pseudospin ferromagnets. What external parameters (displacement field, out of plane magnetic field) are optimal for their realization?
Are there any FCI states in this system that is not generated through the typical vortex attachment procedure, for example non-Abelian states? We believe that future works will be inspired by these questions.
\section*{Acknowledgement}
We thank Daniel Parker, Rahul Sahay, Jie Wang  and Tomohiro Soejima on insightful discussions and helpful comments on the manuscript. A.V. was supported by a Simons Investigator award by the Simons Collaboration on Ultra-Quantum Matter, which is a grant from the Simons Foundation (651440, AV) and by NSF-DMR 2220703. P.J.L. was supported by the Department of Defense (DoD) through the National Defense Science and Engineering Graduate Fellowship (NDSEG) Program. J.Y.L is supported by the Gordon and Betty Moore Foundation under the grant GBMF8690 and by the National Science Foundation under the grant PHY-1748958. \par
\emph{Note added}: During the preparation of the draft we became aware of parallel work by Jie Wang {\it et al.}~\cite{jiewangFCI} which overlaps with part of our results.
\bibliographystyle{apsrev4-1}
\bibliography{references}
\pagebreak
\onecolumngrid
\begin{center}

\textbf{\large Supplementary Information}
\end{center}
\setcounter{equation}{0}
\setcounter{figure}{0}
\setcounter{table}{0}
\makeatletter
\renewcommand{\theequation}{S\arabic{equation}}
\renewcommand{\thefigure}{S\arabic{figure}}
\renewcommand{\thesection}{S\Roman{section}}
\renewcommand{\bibnumfmt}[1]{[S#1]}
\setcounter{section}{0}

\section{Holomorphic Boundary conditions and LLL-like wavefunctions}

In this section we prove a number of results involving ideal band geometry that we claimed in the main text. First we prove the generality of the boundary conditions \eqref{gauge} for a single band with $C>0$.
Then we use a similar argument to Ref. \cite{wangExactLandauLevel2021b}, which assumed a gauge-equivalent version of \eqref{gauge}, to show that all single $C=1$ bands have an LLL form, in the sense of \eqref{eq:ChernOneClassification}.

\subsection{Classification of holomorphic boundary conditions}\label{subsec:classify_bcs}

We begin by showing that for a single ideal band with Chern number $C>0$ it is always possible to perform a gauge transformation $u_k \to \lambda_k u_k$ such that 
\begin{equation} \Xi_\bG(k) = \lambda_{k+G} \lambda^{-1}_k \Xi^\LG_\bG(k-k_0) \label{eq:find_lambda} \end{equation}
where $\bk_0 \simeq \bk_0 + \bG_i/C$ is a vector in the $C^2$-fold reduced first BZ and
\begin{equation}\label{gauge_supp}
  \Xi^\LG_{\bG_1}(k)= \exp(2\pi i C k/G_2), \qquad  \Xi^\LG_{\bG_2}(k)=1   .
\end{equation}
We have added a superscript $\LG$, relative to the main text, to distinguish the choice \eqref{gauge_supp} from other boundary conditions written in this section.

  To our knowledge, \eqref{eq:find_lambda} requires the use of some type of cohomology-based classification; we do not know of an explicit construction.  While the Chern number expression
\begin{equation}
  2\pi C = f_{G_1}(k+G_2) - f_{G_1}(k) + f_{G_2}(k) - f_{G_2}(k+G_1) ,\qquad \Xi_\bG(k) = e^{i f_\bG(k)}
  \label{eq:Chern_from_BCs_supp}
\end{equation}
suggests a linear form of $f(k)$, such that $C$ is manifestly $k$-independent as in \eqref{gauge_supp}, we cannot immediately assume such a linear form of $f_\bG(k)$. Indeed, there are nonlinear functions $f_\bG(k)$ that yield a $k$-independent $C$ in \eqref{eq:Chern_from_BCs_supp}. Consider, for example, applying a nonlinear gauge transformation such as $u_k \to \exp(ik^3/G_1^3)u_k$: such a transformation preserves $C$ but takes a linear $f_\bG(k)$ to a nonlinear $ f_\bG(k)  +3iGk^2/G_1^3$. It is not clear how to directly show that all non-linearities may be gauged away.

Instead, we rely on the relationship of $\Xi_\bG(k)$, often dubbed a ``factor of automorphy'' in the mathematics literature, to holomorphic line bundles \cite{mumford2008abelian,birkenhakeComplexAbelianVarieties2004}. In particular, we consider the set of $\Xi_\bG(k)$ that satisfies the cocycle condition
\begin{equation}
  \Xi_{\bG + \bG'}(k) = \Xi_{\bG'}(k + G) \Xi_\bG(k)
  \label{eq:cocycle_Xi}
\end{equation}
and consider two choices of $\Xi$ equivalent if they are related by a gauge transformation \eqref{eq:find_lambda}. Then a choice of $\Xi$ corresponds exactly to a holomorphic line bundle over the Brillouin Zone torus; the gauge redundancy \eqref{eq:find_lambda} corresponds exactly to the equivalence of line bundles with different choices of transition functions \cite{mumford2008abelian,birkenhakeComplexAbelianVarieties2004}. The product $\Xi_1 \Xi_2$ corresponds to the tensor product of the associated line bundles, and $\Xi \to \Xi^{-1}$ is the mapping to the dual bundle where all transition functions are inverted. The collection of $\Xi$ up to gauge transformations and the set of holomorphic line bundles form isomorphic Abelian groups. 

Holomorphic line bundles on the torus, and therefore boundary conditions $\Xi_\bG(k)$ up to gauge transformations, can be classified via the Appell-Humbert theorem \cite{mumford2008abelian,birkenhakeComplexAbelianVarieties2004}. The Appell-Humbert theorem yields a single canonical choice of $\Xi_\bG(k)$ for each distinct line bundle. This canonical choice is usually written in a ``symmetric gauge'' form rather than our choice \eqref{gauge_supp}, though they may be related through a suitable gauge transformation.

Instead we will use the Appell-Humbert theorem for $C=0$, which in practice is how the whole theorem is proved \cite{mumford2008abelian,birkenhakeComplexAbelianVarieties2004}. The theorem states that $\Xi^\naut$ has $C=0$ in \eqref{eq:Chern_from_BCs_supp} if and only if it is gauge equivalent to $e^{- i \bG \cdot \br_0}$ where $\br_0$ is an element of the ``dual torus''\cite{birkenhakeComplexAbelianVarieties2004}, i.e. the real-space unit cell. Now consider some $\Xi_\bG(\br)$ with Chern number $C$. Then $\Xi^\naut=\Xi_\bG(k)/\Xi^\LG_\bG(k)$ has $C=0$ in \eqref{eq:Chern_from_BCs_supp} and therefore is equivalent to $e^{-i \bG \cdot \br_0}$ for some $\br_0$ by Appell-Humbert. Therefore, we may always apply a gauge transformation such that
\begin{equation}
    \Xi_\bG(k) = e^{-i \bG \cdot \br_0} \Xi_\bG^\LG(k).
\end{equation}

We now show how to convert a choice of $\br_0$ in the real-space unit cell to a choice of $k_0 \equiv k_0 + \bG_i/C$ for $C\neq0$. We perform the gauge transformation
\begin{equation}
  \Xi_\bG(k) \to \lambda_{\bk+\bG} \Xi_\bG(k) \lambda_k^{-1}, \qquad \lambda_k = \exp(i\bG_2 \cdot \br_0  k/G_2) 
  \label{eq:shift_r_to_shift_k_part1}
\end{equation}
which has the effect
\begin{equation}
  \Xi_{\bG_1}(k) = e^{-i \bG_1 \cdot \br_0} \Xi_{\bG_1}^\LG(k) \to \Xi_{\bG_1}^\LG(k-k_0), \qquad \Xi_{\bG_2}(k) = e^{-i \bG_2 \cdot \br_0}  \to 1
  \label{eq:shift_r_to_shift_k_part2}
\end{equation}
with
\begin{equation}
  \bk_0 = -\frac{A_\BZ}{2\pi C} \hat{\bz} \times \br_0.
  \label{eq:shift_r_to_shift_k_part3}
\end{equation}
Here $A_{\BZ} = \hat{\bz} \cdot \bG_1 \times \bG_2$ is the area of the Brillouin Zone.
To simplify the first equation of \eqref{eq:shift_r_to_shift_k_part2} we have written $\br_0 = r_{1,0}\ba_1 +  \br_{2,0} \ba_2$ and identified $-r_{1,0} \bG_2 + r_2 \bG_1 = -\frac{A_\bz}{2\pi} \hat{\bz} \times \br_0$ as a map from the unit cell to the first BZ. The further division by $C$ in \eqref{eq:shift_r_to_shift_k_part3} ensures that $k_0$ is equivalent to $k_0 + G_{1,2}/C$ because $\br_0$ is equivalent to $\br_0 + \ba_{1,2}$.

\subsection{LLL form of $C=1$ ideal band wavefunctions}

We now show how to prove \eqref{eq:ChernOneClassification}, copied below for convenience. This argument was previously given in Ref. \cite{wangExactLandauLevel2021b}; we are repeating it here so that our discussion is self-contained. 
\begin{equation}
    \psi_{\bk \alpha}(\br) = \psi_{\bk - (\bk_0 - \bk^\LLL_0)}(\br) \N_{\alpha}(\br).
    \label{claimofinterest}
\end{equation}
Here $\psi_{\bk \alpha}(\br) = e^{i \bk \cdot \br} u_k(\br)$ is the wavefunction for the ideal $C=1$ band of interest written in the holomorphic gauge corresponding to \eqref{gauge_supp} for some $k_0$. The index $\alpha$ corresponds to a non-positional orbital degree of freedom, for example layer. Similarly, $\psi_\bk^\LLL(\br) = e^{i \bk \cdot \br } u_k^\LLL(\br)$ is the LLL wavefunction written in holomorphic gauge \eqref{gauge_supp} which is constructed explicitly in the next section.

Consider the ratio
\begin{equation}
    \frac{\psi_{\bk \alpha}(\br)}{\psi^\LLL_{\bk - (\bk_0 - \bk_0^\LLL)}(\br)} = e^{-i(\bk_0 - \bk_0^\LLL) \cdot \br} \frac{u_{k \alpha}(\br)}{u^\LLL_{k - (k_0 - k_0^\LLL)}(\br) } \equiv \N_{\alpha k}(\br).
\end{equation}
Here, the final equality defines $\N_{\alpha k}(\br)$, as a holomorphic function which we now show is $k$-independent. Note that $\N_{\alpha k}(\br)$ is a periodic under $k \to k + G_i$ because $\psi_{\bk}$ and $\psi^\LLL_{\bk - (\bk_0 - \bk^\LLL_0)}$ have the same boundary conditions. 

Furthermore, $\N_{\alpha k}(\br)$ can have at most one pole: the wavefunction $u^\LLL_{\bk}(\br)$, for fixed $\br$, only has a single zero in the Brillouin Zone. The lack of extra zeros may be seen by through the explicit construction in the next section or proved on general grounds through the expression \cite{wangExactLandauLevel2021b,vortexability}
\begin{equation}
    2\pi i C = \oint_{\partial \BZ} \ln u_k(\br) dk = 2\pi i N_z.
\end{equation}
where the integral around the boundary of the Brillouin Zone counts the number of zeros $N_z$ of $u_k(\br)$ in the first BZ. The relation to the Chern number is recovered by computing the integral using the boundary conditions of $u$ and recovering \eqref{eq:Chern_from_BCs_supp} \cite{wangExactLandauLevel2021b,vortexability}.

Periodic holomorphic functions with only one pole per unit cell are constant. To see this, note that Cauchy's integral theorem $\oint_{\BZ} N_{k\alpha}(\br) dk$ computes the residue of the pole but this integral vanishes since $N_k$ is periodic. Thus $N_k$ is bounded, but bounded holomorphic functions are constant by Liouville's theorem. We have therefore shown that $N_{k \alpha}(\br)$ is $k$-independent, such that we recover \eqref{claimofinterest}. 

We emphasize that this argument relies on $C=1$; higher Chern bands are not related to each other by $k$-independent factors because their ratio may be an elliptic function with more than one pole per unit cell (with total residue zero). This complexity is difficult to work with, which is one way to view the motivation to decompose higher Chern bands into multiple $C=1$ bands.

\section{Magnetic Bloch states of the Lowest Landau level}

In this section we review the form of magnetic translations and magnetic Bloch states in symmetric electromagnetic gauge. We then construct the LLL wavefunctions with a suitable momentum-space gauge \eqref{gauge} and construct the gauge-dependent magnetic translation form \eqref{eq:magneto_momentum_transl_gauge}.

We take our magnetic field to be a constant value of $B>0$ and choose lattice vectors $\bR_{1,2}$ that enclose $2\pi$ flux and corresponding reciprocal lattice vectors $\bQ_{1,2}$ such that $\bQ_i \cdot \bR_j = 2\pi \delta_{ij}$:
\begin{equation}
 B \hat{\bz} \cdot \bR_1 \times \bR_2  = 2\pi = B^{-1} \hat{\bz} \cdot \bQ_1 \times \bQ_2.
  \label{fluxqtz}
\end{equation}
 For our applications to higher Chern bands, the lattice vectors $\bR$ will describe a $\abs{C}$-fold enlarged real space lattice and the reciprocal lattice vectors $\bQ$ will describe the folded Brillouin Zone.

The Hamiltonian is
\begin{equation}
    H = \frac{\abs{-i \bnabla - e \bA}^2}{2m} = \frac{2}{m} \Pi \ov{\Pi} + \frac{B}{2m}, \quad \ov{\Pi} = (-i \ov{\partial} -  \ov{A}) 
\end{equation}
where $2\ov{\partial} = \partial_x + i \partial_y$ and $2 \ov{A} = A_x + i A_y$. The lowest Landau level is the zero mode space of $\ov{\Pi}$. The zero mode wavefunctions are of the form 
\begin{equation}
  \psi(\br) = f(z) e^{-K(\br)}
  \label{inhomWFs}
\end{equation}
where $-i\ov{\partial} K = \ov{A}$.
We will choose the symmetric gauge vector potential $\bA = B(-y,x)$ such that $2\ov{A} = B(-y + ix) = -iBz$ and $K = Bz\ov{z}/4$.

The magnetic translation operators are given by
\begin{equation}
    T^\magt_{\bl}= e^{-i\zeta_{\bl} (\br)}e^{\bl \cdot \bnabla},
\end{equation}
where
\begin{equation}
    \bA(\br + \bl) - \bA(\br) = \bnabla \zeta_\bl(\br)
\end{equation}
From this we can derive the commutation relation
\begin{equation}
    T^\magt_{\bl_1} T^\magt_{\bl_2} = e^{ i B\hat{z} \cdot \bl_1 \times \bl_2}T^\magt_{\bl_2} T^\magt_{\bl_1} 
    \label{eq:magnetic_transl_commutator}
\end{equation}
such that the magnetic translations commute if $\bl_i$ are lattice vectors $\bR$ due to \eqref{fluxqtz}. Note, however, that there is a subtlety $T^\magt_{\bR_1} T^\magt_{\bR_2} = - T^\magt_{\bR_1 + \bR_2}$ since the triangle spanned by the points $\bR_1, \bR_2, \bR_1 + \bR_2$ encloses $\pi$ flux. We therefore take $(T^\magt_{\bR_1})^m (T^\magt_{\bR_2})^n $ to be the Bloch translation associated to the lattice vector $m \bR_1 + n \bR_2$, which in general is not the same as $T^\magt_{m \bR_1 + n \bR_2}$.

Consider Bloch wavefunctions $\psi_\bk(\br)$ such that
\begin{equation}
  T^\mBloch_{\bR} \psi_\bk(\br) = e^{i \bk \cdot \bR} \psi_\bk(\br).
  \label{magneticbloch}
\end{equation}
On general grounds, using \eqref{eq:magnetic_transl_commutator} we have
\begin{equation}
    T^\mBloch_{\bR} T^\magt_{\bl} \psi_\bk = e^{iB \hat{\bz} \cdot \bR \times \bl} T^\magt_{\bl} T^\mBloch_{\bR} \psi_\bk = e^{iB \hat{\bz} \cdot \bR \times \bl} e^{i \bk \cdot \bR} T^\magt_{\bl} \psi_\bk = e^{i(\bk - B\hat{\bz} \times \bl)\cdot \bR } e^{i \bk \cdot \bR} T^\magt_{\bl}\psi_\bk
\end{equation}
such that $T^\magt_\bl \psi_\bk$ and $\psi_{\bk - B\hat{\bz} \times \bl}$ have the same eigenvalue under translations. We therefore conclude that $T^\magt_\bl$ acts as a momentum translation
\begin{equation}
    T^\magt_\bl \psi_{\bk} \propto \psi_{\bk - B\hat{\bz} \times \bl},
    \label{magnetic_transl_general_gauge_supp}
\end{equation}
where the proportionality constant is gauge dependent because the relative prefactor between $\psi_\bk$ and $\psi_{\bk - B \hat{\bz} \times \bl}$ is gauge dependent. We now move onto gauge-specific calculations, in both the electromagnetic and momentum-space sense.

In our symmetric gauge, we have the translation operators
\begin{equation}
    T^\magt_{\bl} = e^{-\frac{i}{2}B \bl \times \br} e^{\bl \cdot \bnabla}, \qquad T^\mBloch_\bR = \eta_\bR T^\magt_{\bR}
\end{equation}
and the wavefunctions \cite{wangChiralApproximationTwisted2021a,wangExactLandauLevel2021b}
\begin{equation}
  \psi_\bk(\br) = \lambda_k e^{\frac{i}{2} \ov{k}' z} \sigma(z + iB^{-1} k') e^{-B\abs{z}^2/4} 
  \label{magneticblochWF}
\end{equation}
where $k' = k - k_0^\LLL$ with $k_0^\LLL =-(\bQ_1 + \bQ_2)/2$ and $\sigma(z) = \sigma(z | R_1, R_2)$ is the (modified \cite{haldaneModularinvariantModifiedWeierstrass2018}) Weierstrass sigma function which satisfies
\begin{equation}
\sigma(-z) = - \sigma(z), \qquad 
\sigma(z + R) = \eta_\bR e^{\frac{B}{2}\ov{R}\left(z + \frac{R}{2}\right)}
\label{eq:weier_translation_properties}
\end{equation}
where $\eta_\bR = 1$ if $\bR/2$ is a lattice vector and $-1$ otherwise.
From \eqref{eq:weier_translation_properties} it is straightforward to verify the Bloch boundary conditions of \eqref{magneticblochWF}. The relations $\hat{\bz} \cdot \bR \times \br = -\frac{i}{2}(\ov{R} z - R \ov{z})$ and $\bR \cdot \br = \frac{1}{2}(\ov{R} z + R \ov{z})$ for transforming between vector and complex notation are helpful here. 

The wavefunction \eqref{magneticblochWF} is in a holomorphic gauge; this can be straightforwardly verified by computing $u_k(\br) = e^{-i \bk \cdot \br} \psi_\bk(\br) = e^{-\frac{i}{2} \ov{k} z} e^{-\frac{i}{2} z \ov{k}} \psi_\bk(\br)$ such that the non-holomorphic $e^{\frac{i}{2} \ov{k} z}$ factor in $\psi$ is cancelled. The holomorphic factor $\lambda_k$ parameterizes the remaining holomorphic gauge freedom; we now fix $\lambda_k$ such that the boundary conditions of \eqref{magneticblochWF} match \eqref{gauge}, \eqref{gauge_supp} for $k_0=0$. To do this, we simply engineer $\lambda_k$ such that $\Xi_{\bQ_2}(k) = 1$. We obtain
\begin{equation}
    1 = \lambda_{k+Q_2}\lambda_k^{-1} e^{i\pi + \frac{1}{2B}\ov{Q}_2\left(k'+\frac{Q_2}{2}\right)} \implies \lambda_k = \exp\left(-i\pi\frac{k'}{Q_2}-\frac{\ov{Q}_2k'^2}{4BQ_2} \right).
    \label{eq:find_lambda_LLL}
\end{equation}
We now compute, using \eqref{fluxqtz}
\begin{equation}
    \Xi_{\bQ_1}(k) = \lambda_{k+Q_1} \lambda_k^{-1}e^{i\pi + \frac{1}{2B}\ov{Q}_1\left(k'+\frac{Q_1}{2}\right)} 
    = e^{-i\pi \frac{Q_1}{Q_2}} \exp\left(\frac{1}{2} B^{-1}(\ov{Q_1} Q_2 - Q_1 \ov{Q_2})\left(k'+\frac{Q_1}{2}\right)/Q_2 \right)
    = e^{2\pi i k'/Q_2}
\end{equation}
such that we obtain the desired boundary conditions \eqref{gauge_supp} with $k_0 = k_0^\LLL = -(Q_1 + Q_2)/2$.

We now compute $c_\bl(k)$ where
\begin{equation}
    T^\magt_\bl \psi_\bk = c_\bl(\bk) \psi_{\bk - B \hat{\bz} \times \bl }.
\end{equation}
Through direct computation we obtain
\begin{equation}
\begin{aligned}
    c_\bl(\bk) & = \frac{\lambda_k}{\lambda_{k-iBl}} \exp\left[\frac{i}{2} l\left(\ov{k}' - \frac{i B \ov{l}}{2}\right) \right]\\
    & = e^{\pi B l/Q_2} \exp \left[ \frac{-i}{2} \frac{\ov{Q}_2}{Q_2} l\left(k'-\frac{iBl}{2}\right) + \frac{i}{2} l \left(\ov{k}' + \frac{iB\ov{l}}{2} \right) \right]\\
    & = e^{i \bk \cdot \bl} e^{\frac{i}{2} (\bQ_1 + \bQ_2) \cdot \bl} e^{\pi B l/Q_2} \exp \left[ - \frac{i}{2} \left(\frac{\ov{Q}_2}{Q_2}l + \ov{l} \right)\left(k' - \frac{iBl}{2}\right) \right] \\
    & = e^{i \bk_0^\LLL} c_{\bl}^\naut(\bk - \bk_0^\LLL)
   \end{aligned} 
\end{equation}
where
\begin{equation}
    c^\naut_\bl(\bk)  = 
    e^{i \bk \cdot \bl} e^{\frac{i}{2}(\bQ_1 + \bQ_2) \cdot \bl}e^{\frac{\pi B l}{Q_2} - \frac{i}{2} \left(\frac{\ov{Q}_2}{Q_2}l + \ov{l} \right)\left(k - \frac{iBl}{2}\right) } 
\end{equation}
as claimed in the main text \eqref{eq:magneto_momentum_transl_gauge}.

\section{Computation details of $U_k$}\label{Uk}

In this section we construct and prove the uniqueness of the matrix $U_k$ \eqref{Uk} that hybridizes the folded ideal $C=2$ band into two ideal $C=1$ bands. 
We find it easier to compute the wavefunction of the ideal bands after decomposition directly, and then organize the coefficients into the matrix $U_k$.\par
We take a general holomorphic combination
\begin{equation}
    \phi_\bk=\alpha_0(k)\psi_\bk+\alpha_1(k)\psi_{\bk+\bG_1/2}
    \label{eq:holo_ansatz}
\end{equation}
and require that
\begin{equation}
    \phi_{\bk+\bG_1/2}=\xi_{1,g}(k)\phi_\bk, \qquad \phi_{\bk + \bG_2} = \phi_\bk
    \label{eq:demanded_bcs}
\end{equation}
We will find two distinct solutions $\alpha_0, \alpha_1$; these two solution sets will comprise the two rows of $U_k$. 
Inserting \eqref{eq:holo_ansatz} into \eqref{eq:demanded_bcs}, we obtain the following conditions for $\alpha_{0,1}$:
\begin{equation}\label{coeffrel}
    \begin{aligned}
        &\alpha_0(k+G_1/2)=\alpha_1(k)\xi_{1,g}(k)\\
        &\alpha_1(k+G_1/2)\xi_{2,k_0}(k)=\alpha_0(k)\xi_{1,g}(k)
    \end{aligned}
\end{equation}
The first equation gives $\alpha_1(k)=\alpha_0(k+G_1/2)\xi^{-1}_{1,g}(k)$; further using the second equation, we have
\begin{equation}
    \alpha_0(k+G_1)\xi^{-1}_{1,g}(k+G_1/2)\xi_{2,k_0}(k)=\alpha_0(k)\xi_{1,g}(k),
\end{equation}
which means that
\begin{equation}
    \frac{\alpha_0(k+G_1)}{\alpha_0(k)}=\exp(2\pi i (2k_0-2g+G_1/2)/G_2)=\exp(\beta).
    \label{eq:betadefn}
\end{equation}
for some $k$-independent $\beta$. We also have $\alpha(k + G_2) = \alpha(k)$ from the second equation in \eqref{eq:demanded_bcs}.
Defining $\alpha_0(k)=e^{\gamma(k)}$ then yields
\begin{equation}
    \gamma(k+G_1)-\gamma(k)=\beta, \qquad \gamma(k + G_2) - \gamma(k) = 2\pi i n.
\end{equation}
Since $\gamma(k)$ has no poles, a contour integral along boundaries of the Brillouin zone yields, via the residue theorem,
\begin{equation}
    \beta G_2 - 2\pi i n G_1 = 0.
\end{equation}
We multiply the wavefunction by an overall constant such that $\alpha_0(k=0) = 1$ and $\gamma(0)=0$. We define $\delta(k)=\gamma(k)-\beta k/G_1$ such that $\delta(k)$ vanishes on reciprocal lattice vectors $a G_1+bG_2$ for $a,b\in\mathbb{Z}$. Since $\delta(k)$ is a periodic holomorphic function it must be constant, such that $\delta(0)=0$, $\delta(k)=0$ and thus $\gamma(k)=2\pi i n k/G_2$.\par
From \eqref{eq:betadefn}, we deduce that for each $g$ that satisfies $\exp(2\pi i (2k_0-2g+G_1/2 + n G_1)/G_2)$
\eqref{eq:demanded_bcs}. 
Taking a logarithm, and encoding the associated $2\pi i$ ambiguity with an integer $m$, we obtain
\begin{equation}
    g=k_0+G_1/4-n G_1/2-mG_2/2.
\end{equation}
The shift in $G_1/2$ may be gauged away because $\bG_1/2$ is now a reciprocal lattice vector (see SI Sec. \ref{subsec:classify_bcs}). We can then choose $n=0$ such that $\alpha_0 = 1$. Since $\bG_2$ is a reciprocal lattice vector, there are two distinct choices are $m=0,1$. We then obtain the two possible boundary conditions 
\begin{equation}
    \xi_m=\xi_{1,k_0+G_1/4-mG_2/2}(k)=(-1)^m\xi_{1,k_0+G_1/4}(k)
\end{equation}
for which the wavefunctions $\phi_{m\bk}$ are
\begin{equation}
    \phi_{m\bk}=\psi_\bk+(-1)^m\xi_{1,k_0+G_1/4}^{-1}(k)\psi_{\bk+\bG_1/2}.
\end{equation}
We may then read off the matrix $U_k$ \begin{equation}
    U_k=\left(\begin{array}{cc}
        1 & \xi_{1,k_0+G_1/4}^{-1}(k) \\
        1 & \xi_{1,k_0+G_1/4-G_2/2}^{-1}(k)
    \end{array}\right).
\end{equation}
\section{Chern Bands with $C>2$}
For $C>2$ bands, a similar procedure can be performed. However, in general we can attempt to generate $r$ bands, in which $r<C$. We break translation symmetry in $\ba_1$ direction by $r$ times:
\begin{equation}
    \phi_\bk=\sum_m\alpha_m(k)\psi_{\bk+m\bG_1/r}
    \label{eq:higherChernhybridize}
\end{equation}
and try to ``diagonalize'' the boundary conditions:
\begin{equation}
    \phi_{\bk+\bG_1/r}=\xi_{1,g}(k)\phi_\bk, \qquad \phi_{\bk + \bG_2} = \phi_\bk
    \label{eq:higherChern_solveBCs}
\end{equation}
We find that when $r$ doesn't divide $C$ the two boundary conditions \eqref{eq:higherChern_solveBCs} are not compatible. When $r$ does divide $C$ we find $r$ solutions to \eqref{eq:higherChern_solveBCs}, leading to an $r \times r$ holomorphic matrix $U_k$. The resulting bands \eqref{eq:higherChernhybridize} have Chern number $C/r$. We must solve the following equations similar to \eqref{coeffrel}:
\begin{equation}\label{coeffrelHC}
    \begin{aligned}
        &\alpha_0(k+G_1/r)=\alpha_1(k)\xi_{C/r,g}(k)\\
        &\alpha_1(k+G_1/r)=\alpha_2(k)\xi_{C/r,g}(k)\\
        &\dots\\
        &\alpha_{r-1}(k+G_1/r)\xi_{C,k_0}(k)=\alpha_0(k)\xi_{C/r,g}(k)
    \end{aligned}
\end{equation}
All coefficients are fundamentally related to $\alpha_0$. Using techniques analogous to those in the previous section, we find
\begin{equation}
    U_{ml}=\alpha_{ml}(k)=e^{-2\pi i ml/r}\prod_{n<l}\xi^{-1}_{C/r,k_0+(r-1)G_1/2r}(k+nG_1/r)
\end{equation}
We get $r$ bands $\phi_{m\bk}$, each with Chern number $C/r$, with boundary conditions 
\begin{equation}
    \phi_{m (\bk + \bG_1/r)} = \xi_{C/r,k_0+(r-1)G_1/2r}(k) e^{2\pi i m/r} \phi_{m\bk}, \qquad \phi_{m (\bk+\bG_2)} = \phi_{m\bk}
\end{equation}

We now specialize to the case $r=C$ in order to derive the $\SU(C)$ action.
We find that $\phi_{m(\bk+\bG_2/2-m\bG_2/C)}$ all have the same boundary conditions. We thus define
\begin{equation}
    \tilde{\phi}_{m\bk}=\phi_{m(\bk+\bG_2/2-m\bG_2/C)}
\end{equation}

Similar to $C=2$, these states also form a non-unitary representation of $\SU(C)$~\cite{FAIRLIE_SUC}, with the similar identification
\begin{equation}
\begin{aligned}
    \mathsf{Z}&=(T^\magt_{\ba_2})^{-1}T_{\ba_2},\\ \mathsf{X}&=(T^\magt_{\ba_1})^{-1}T_{\ba_1}, \,\, \mathsf{Z}\mathsf{X}=e^{2\pi i/C}\mathsf{X}\mathsf{Z}
\end{aligned}
\end{equation}
Next, we want to understand the ground state degeneracy in this system. In order to obtain the degeneracy of the $\SU(C)$ ferromagnet in a finite system, we need to identify the $\SU(C)$ irreducible representation for the ground states. One definite element of the ground state manifold is the state with full-polarization in a certain $C=1$ band: $\ket{0}=\mathcal{A}\ket{\phi_{0\bk_1},\phi_{0\bk_2},\dots}$ in which $\mathcal{A}$ antisymmetrizes particles. Since it is fully symmetrized with respect to the $SU(C)$ index, it is part of the irrep expressed by the following Young diagram:
\begin{equation}
  \Yvcentermath1 \underbrace{ {\yng(2) \cdots\cdots \yng(1)} }_{N_e}
\end{equation} 
whose dimension can be calculated by the Hook's formula:
\begin{equation}\label{hook}
    \textrm{dim\,[$N_e$]$_C$} = \frac{(N_e+C-1)!}{N_e! (C-1)!}
\end{equation}
As the state $\ket{0}$ will generate all other state in the same irrep under the action of $\SU(C)$ symmetry, this corresponds to the groundstate degeneracy of the system. For $C=2$, $\textrm{GSD} = N_e+1$. 

As with $C=2$, because the $\SU(C)$ operators involve momentum translations by $\bG_{i}/C$ the full $\SU(C)$ degeneracy only appears in finite size $N_1 \ba_1 \times N_2 \ba_2$ systems when both $N_1$ and $N_2$ are divisible by $C$.
If only $N_1$ or $N_2$ are divisible by $C$, but not the other, there are only $C$ ground states, each one obtained by completely filling one of the $\phi_{m\bk}$ bands. 
\section{General Unit Cell Expansion}
In this section we examine the possibility of expanding the unit cell by a factor of $r$, but not necessarily entirely along the $\ba_1$ or $\ba_2$ direction. We will construct $r$ bands with Chern number $C/r$, though for the $\SU(C)$ action one should follow the previous procedure and expand the unit cell by a factor of $C$ in both directions.

We call our original lattice vectors $\ba_1,\ba_2$ and the new lattice vectors $\ba'_1,\ba'_2$. The original reciprocal lattice vectors are $\bG_1,\bG_2$ and the new ones $\bG'_1,\bG'_2$. We write down the coordinates of $\ba'$ in terms of $\ba$:
\begin{equation}
    \ba'_i=\sum_j M_{ij}\ba_j
\end{equation}
where $M_{ij}$ are integer. $\det M$ is the ratio between the volume of the two unit cells: that is, $\det M=r$. We take $M$ to have positive determinant to preserve orientation.\par
We consider the Smith normal form of the matrix $M$:
\begin{equation}
    M=PRQ,\quad R=\left(\begin{array}{cc}
        r_1 & 0 \\
        0 & r_2
    \end{array}\right)
\end{equation}
in which $P,Q\in SL(2,\mathbb{Z})$ and $r_1$ divides $r_2$. Physically speaking, it tells us that any expansion of unit cell can be decomposed into three steps: redefining the unit cell in the original lattice, expanding the new cell by $r_1\times r_2$ times and redefining the unit cell again. Lattice redefinitions do not affect the Chern number or holomorphicity of bands $\phi_{mk}$, so without loss of generality we may set $P=Q=1$ by choosing an appropriate basis for the original and final unit cell.\par
Note that $\det R = r=r_1 r_2$. We will prove that in general such an unit cell expansion for an ideal Chern number $C$ band will generate $r$ ideal bands, each with Chern number $C/r$.\par
We have $\bG'_1=\bG_1/r_1$, $\bG'_2=\bG_2/r_2$. We define the wavefunctions in the folded Brillouin zone as
\begin{equation}
    \psi_{lm\mathbf{k}}=\psi_{l\bG_1'+m\bG_2'+\bk},
\end{equation}
which have the boundary conditions
\begin{equation}
\begin{aligned}
        \left(\Xi_{\bG_1'}(\bk)\right)_{lm,l'm'}&=\delta_{mm'}\delta_{l,l'+1} \text{ if }1\leq l'\leq r_1,\\ \left(\Xi_{\bG_1'}(\bk)\right)_{r_1m,0m'}&=\delta_{mm'}\xi_{C,k_0}(\bk+m\bG_2')
\end{aligned}
\end{equation}
and
\begin{equation}
    \left(\Xi_{\bG_2'}(\bk)\right)_{lm,l'm'}=\delta_{ll'}\delta_{m,m'+1 \text{mod }n_2}.
\end{equation}
The ``eigenfunctions'' after we diagonalize both $\Xi_{\bG_1'}$ and $\Xi_{\bG_2'}$ are
\begin{equation}
    \varphi_{\alpha\beta\bk}=\sum_{lm}\exp\left(-2\pi i \left(\frac{\alpha l}{r_1}+\frac{\beta m}{r_2}\right)\right)\prod_{n<l}\xi^{-1}_{C/r_1,k_0+(r_1-1)G_1/2r_1}(k+nG_1')\psi_{lm\bk}
\end{equation}
for $\alpha \equiv \alpha + r_1$ and $\beta \equiv \beta + r_2$. We therefore obtain $r_1 r_2 = r$ distinct bands in total as expected.
The boundary conditions for $\varphi_{\alpha \beta \bk}$ are
\begin{equation}
    \xi_{\bG_1\alpha\beta}(k)=\exp(2\pi i\frac{\alpha}{r_1})\xi_{C/r_1,k_0+(r_1-1)G_1/2r_1}(k),
        \label{eq:generalfoldBCs}
\end{equation}
and 
\begin{equation}
    \xi_{\bG_2\alpha\beta}(k)=\exp(2\pi i\frac{\beta}{r_2})    
\end{equation}
to convert into the canonical form, we define
\begin{equation}
    \phi_{\alpha\beta\bk}=\varphi_{\alpha\beta\bk}\exp(-2\pi i\beta k/G_2)
\end{equation}
These $\phi$ states are holomorphic because they are holomorphic combinations of ideal bands. Furthermore, they have Chern number $C/r$ as may be identified from the boundary condition \eqref{eq:generalfoldBCs} and computed through \eqref{Chern_from_BCs}.

\end{document}